\documentclass[conferemce]{IEEEtran}
\usepackage{graphicx}
\usepackage{epstopdf}
\usepackage{amsthm}
\usepackage{epsfig}
\usepackage{latexsym}
\usepackage{amsfonts}
\usepackage{here}
\usepackage{rawfonts}
\usepackage[latin1]{inputenc}
\usepackage[T1]{fontenc}
\usepackage{calc}
\usepackage{capitalgreekitalic}
\usepackage{url}
\usepackage{enumerate}
\usepackage{color}
\usepackage{url}
\usepackage[tbtags]{amsmath}
\usepackage{amssymb}
\usepackage{upref}
\usepackage{epic,eepic}
\usepackage{times}
\usepackage{dsfont}
\usepackage{comment}
\usepackage{cite}
\usepackage{bbm}
\usepackage{mathrsfs}
\usepackage{bm}
\makeatletter
\newcommand*\bigcdot{\mathpalette\bigcdot@{.5}}
\newcommand*\bigcdot@[2]{\mathbin{\vcenter{\hbox{\scalebox{#2}{$\m@th#1\bullet$}}}}}
\makeatother

\newtheorem{theorem}{\bf Theorem}

\newtheorem{lemma}{\bf Lemma}

\newtheorem{definition}{\bf Definition}

\usepackage{dsfont}

\newcounter{step}
\newlength{\totlinewidth}
\newenvironment{algorithm}{%
  \rule{\linewidth}{1pt}
  \begin{list}{}%
    {\usecounter{step}%
      \settowidth{\labelwidth}{\textbf{Step 2:}}%
      \setlength{\leftmargin}{\labelwidth}%
      \setlength{\topsep}{-2pt}%
      \addtolength{\leftmargin}{\labelsep}%
      \addtolength{\leftmargin}{2mm}%
      \setlength{\rightmargin}{2mm}%
      \setlength{\totlinewidth}{\linewidth}%
      \addtolength{\totlinewidth}{\leftmargin}%
      \addtolength{\totlinewidth}{\rightmargin}%
      \setlength{\parsep}{0mm}%
      \raggedright}}%
  {\end{list}%
  \rule{\linewidth}{1pt}}
\newcounter{substep}

  {\end{list}}

\newlength{\aligntop}
\newlength{\alignbot}
 \makeatletter
\renewenvironment{align}{%
  \vspace{\aligntop}
  \start@align\@ne\st@rredfalse\m@ne
}{%
  \math@cr \black@\totwidth@
  \egroup
  \ifingather@
    \restorealignstate@
    \egroup
    \nonumber
    \ifnum0=`{\fi\iffalse}\fi
  \else
    $$%
  \fi
  \ignorespacesafterend%
  \vspace{\alignbot}\par\noindent
} \makeatother

\IEEEoverridecommandlockouts

\usepackage{algorithm}
\usepackage{algorithmic}
\usepackage{array}

\usepackage{makeidx}
\makeindex

\makeatletter
\newcommand\semihuge{\@setfontsize\semihuge{19.3}{25}}
\makeatother

\makeatletter
\newcommand\semismall{\@setfontsize\semihuge{12.4}{15}}
\makeatother

\allowdisplaybreaks[4]
\usepackage{subfigure}

\begin{document}
\title{\huge Distributed Reinforcement Learning for Age of Information Minimization in Real-Time IoT Systems}

\author{{Sihua Wang,} \emph{Student Member, IEEE}, {Mingzhe Chen,} \emph{Member, IEEE},{ Zhaohui Yang}, \emph{Member, IEEE},\\
{Changchuan Yin}, \emph{Senior Member, IEEE}, Walid Saad, \emph{Fellow, IEEE}, \\
{Shuguang Cui}, \emph{Fellow}, \emph{IEEE}, and {H. Vincent Poor}, \emph{Fellow}, \emph{IEEE}\\
\thanks{\scriptsize S. Wang and C. Yin are with the Beijing Laboratory of Advanced Information Network, and the Beijing Key Laboratory of Network System Architecture and Convergence, Beijing University of Posts and Telecommunications, Beijing 100876, China. Emails: \protect\url{sihuawang@bupt.edu.cn;} ccyin@ieee.org.}
\thanks{\scriptsize M. Chen is with the Department of Electrical and Computer Engineering, Princeton University, Princeton, NJ, 08544, USA, Email: \protect\url{mingzhec@princeton.edu}.}
\thanks{\scriptsize Zhaohui Yang is with the Department of Engineering, King's College London, WC2R 2LS, UK, Email: \protect\url{yang.zhaohui@kcl.ac.uk}.}
\thanks{\scriptsize W. Saad is with the Wireless@VT, Bradley Department of Electrical and Computer Engineering, Virginia Tech, Blacksburg, VA, 24060, USA, Email: \protect\url{walids@vt.edu}.}
\thanks{\scriptsize S. Cui is with the Shenzhen Research Institute of Big Data (SRIBD) and the Future Network of Intelligence Institute (FNii), Chinese University of Hong Kong, Shenzhen, 518172, China, Email: \protect\url{robert.cui@gmail.com}.}
\thanks{\scriptsize H. V. Poor is with the Department of Electrical and Computer Engineering, Princeton University, Princeton, NJ, 08544, USA, Email: \protect\url{poor@princeton.edu}.}
}

\vspace{-1.8cm}

\maketitle
\pagestyle{empty}  
\thispagestyle{empty} 

\begin{abstract}
In this paper, the problem of minimizing the weighted sum of age of information (AoI) and total energy consumption of Internet of Things (IoT) devices is studied. In the considered model, each IoT device monitors a physical process that follows nonlinear dynamics. As the dynamics of the physical process vary
over time, each device must find an optimal sampling frequency to sample the real-time dynamics of the physical system and send sampled information to a base station (BS). Due to limited wireless resources, the BS can only select a subset of devices to transmit their sampled information. Thus, edge devices must cooperatively sample their monitored dynamics based on the local observations and the BS must collect the sampled information from the devices immediately, hence avoiding the additional time and energy used for sampling and information transmission. To this end, it is necessary to jointly optimize the sampling policy of each device and the device selection scheme of the BS so as to accurately monitor the dynamics of the physical process using minimum energy. This problem is formulated as an optimization problem whose goal is to minimize the weighted sum of AoI cost and energy consumption. To solve this problem, we propose a novel distributed reinforcement learning (RL) approach for the sampling policy optimization. The proposed algorithm enables edge devices to cooperatively find the global optimal sampling policy using their own local observations. Given the sampling policy, the device selection scheme can be optimized thus minimizing the weighted sum of AoI and energy consumption of all devices. Simulations with real data of PM 2.5 pollution show that the proposed algorithm can reduce the sum of AoI by up to 17.8\% and 33.9\% and the total energy consumption by up to 13.2\% and 35.1\%, compared to a conventional deep Q network method and a uniform sampling policy.

\end{abstract}



%
\IEEEpeerreviewmaketitle

\section{Introduction}
For Internet of Things (IoT) applications such as environmental monitoring and vehicle tracking, the freshness of the status information of the physical process at the devices is of fundamental importance for accurate monitoring and control. To quantify the freshness of the status information of sensor data, age of information (AoI) has been proposed as a performance metric \cite{WLAOI}. AoI is defined as the duration between the current time and the generation time of the most recently received status update. Compared to conventional delay metrics that measure queuing or transmission latency, AoI considers the generation time of each measurement, thus characterizing the freshness of the status information from the perspective of the destination. Therefore, optimizing AoI in IoT leads to distinctively different system designs from those used for conventional delay optimization.

\subsection{Related Works}

The existing literatures such as in [2]--[7] focused on many key AoI problems in IoT settings. In particular, in \cite{RDY}, the authors optimized wireless resource allocation to minimize the average instantaneous AoI. The authors in \cite{XWJY} derived the analytical expression of the average AoI for different sensors in a computing-enabled IoT system. In \cite{XSZZ}, an uplink grant-free massive access protocol is introduced for an IoT network with multiple channels to minimize the sum AoI of the IoT devices. The authors in \cite{SL1} optimized the AoI of each user under a sampling cost constraint. The AoI of each energy harvesting transmitter is minimized for both first-come first-serve and last-come first-serve systems in \cite{SL2}. The authors in \cite{SFJY} designed an age-oriented relaying protocol to minimize the average AoI of IoT devices. However, the existing works in \cite{RDY}--[7] only investigated the optimization of the sampling policy without considering the dynamics of the physical process. In fact, the dynamics of a realistic physical process in a cyber-physical system such as the IoT will strongly influence the optimization of the sampling policy of each device and the device selection scheme of the base station (BS). For example, as the physical process varies rapidly, an IoT device must increase its sampling frequency to capture these physical dynamics. Meanwhile, the BS must immediately allocate the limited wireless resources to those devices that have a higher frequency for physical dynamics transmission. In contrast, as the physical process slowly varies, an IoT device can save energy by reducing its sampling frequency. Therefore, the analysis of the real-world dynamics in each physical process will seriously affect the optimization of the sampling and transmission schemes. However, since the dynamics of the monitored physical process are not available for the BS until the devices sample the physical process and upload their sampled information to the BS successfully, the BS may not be able to find the optimal sampling and transmission schemes using the traditional optimization methods in \cite{RDY}--[7]. To address this challenge, one promising solution is to use reinforcement learning (RL) to allow the BS to estimate the dynamics of monitored physical process and find the optimal sampling and transmission schemes.



Recently, a number of existing works such as in \cite{MAHS}--[12] used RL algorithms to solve problems involving AoI as a performance measure. In \cite{MAHS}, the authors developed an RL algorithm for optimizing resource allocation so as to minimize the sum of AoI of all source nodes. The authors in \cite{HBBEU} used RL methods to make scheduling decisions that are resilient to network conditions and packet arrival processes. In \cite{MVW}, the authors optimized the caching content update scheme to minimize the long-term average AoI of users in a heterogeneous network. The authors in \cite{AHMM} proposed a low complexity RL algorithm to minimize the sum of the expected AoI of all sensors. In \cite{AMWH}, the authors studied the use of a new RL framework to optimize the AoI in a drone-assisted wireless network. However, most of these works \cite{MAHS}--[12] used centralized RL algorithms to determine the sampling and transmission decisions of all devices. In such centralized scenarios, each edge device can only sample the physical process after receiving the sampling command from the BS, which incurs an additional delay for environment monitoring and control. Moreover, using centralized RL algorithms, the BS must update the sampling and transmission schemes based on the entire set of the devices' local observations and actions whose dimension increases exponentially with the number of devices. To address these challenges, one can use distributed RL solutions allowing each device to train its own machine learning model so as to determine the sampling action immediately. The authors in \cite{MCHX} proposed the use of a distributed RL to minimize the energy used to transmit the sampled information  under the AoI constraint. In \cite{JHLRH}, the authors developed a distributed deep RL algorithm to optimize device-to-device packet delivery over limited spectrum resources. The authors in \cite{JHKLZ} used a distributed sense-and-send protocol to minimize the average AoI. However, using the distributed algorithms in \cite{MCHX}--[15], each device can only train its local learning model with the local observation of the monitored physical process. Therefore, IoT devices may not be able to find an optimal sampling and transmission schemes. In consequence, it is necessary to develop a novel distributed RL algorithm allowing IoT devices to cooperatively update the RL parameters based on the individual observation of each device, thus finding the optimal sampling and transmission schemes.




\subsection{Contributions}

The main contribution of this paper is a novel framework that enables a BS and devices in an IoT system to cooperatively monitor realistic physical processes simultaneously with a minimum AoI cost and energy consumption. Our key contributions include:
\begin{itemize}
\item We consider a real-time IoT system in which cellular-connected wireless IoT devices transmit their sampled information of numerous monitored physical processes to a BS that captures the dynamics of each physical process. For the considered model, the impact of the actual dynamics of each physical process on the sampling frequency of each device is explicitly considered. In addition, the wireless resources used for dynamic process transmission are limited and, hence, the BS needs to select an appropriate subset of devices to upload their status packets so as to reconstruct the  monitored physical processes accurately.


\item For this purpose, we first derive a closed-form expression for the relationship between the dynamics of the physical process and the sampling frequency of each device. Based on this relationship, the BS and IoT devices can cooperatively adjust the dynamic process sampling and uploading scheme so as to enable the BS to accurately reconstruct the monitored physical process. This joint sampling and device selection problem is formulated as an optimization problem whose goal is to minimize the weighted sum of AoI and energy consumption of all devices.


\item To solve this optimization problem, a distributed QMIX algorithm is proposed to find the global optimal sampling policy for the devices. Compared to traditional RL algorithms, the proposed method enables each device to use its local observation to estimate the Q-value under global observation. Thus, with the proposed distributed QMIX algorithm, devices can find the optimal sampling policy using their local observations and yield a better performance compared to the one achieved in \cite{MCHX}--[15]. Given the sampling policies of all devices, the BS can directly optimize the device selection scheme using dynamic programming.


\end{itemize}

Simulations with real data of PM 2.5 pollution show that, compared to the conventional deep Q-network (DQN) method and the uniform sampling policy, the proposed algorithm can reduce the sum of AoI by up to 17.8\% and 33.9\% and the total energy consumption by up to 13.2\% and 35.1\%, respectively. \emph{To the best of our knowledge, this is the first work that considers the optimization of the sampling policy and device selection scheme for a real-time IoT system that consists of numerous realistic physical processes.}


The rest of this paper is organized as follows. The system model and the problem formulation are described in Section II. Section III discusses the proposed learning framework for the optimization of the sampling policy and device selection scheme. In Section IV, numerical results are presented and discussed. Finally, conclusions are drawn in Section V.

\section{System Model and Problem Formulation}\label{se:system}
\begin{figure}[t]
\centering
\setlength{\belowcaptionskip}{-0.45cm}
\vspace{-0.1cm}
\includegraphics[width=9cm]{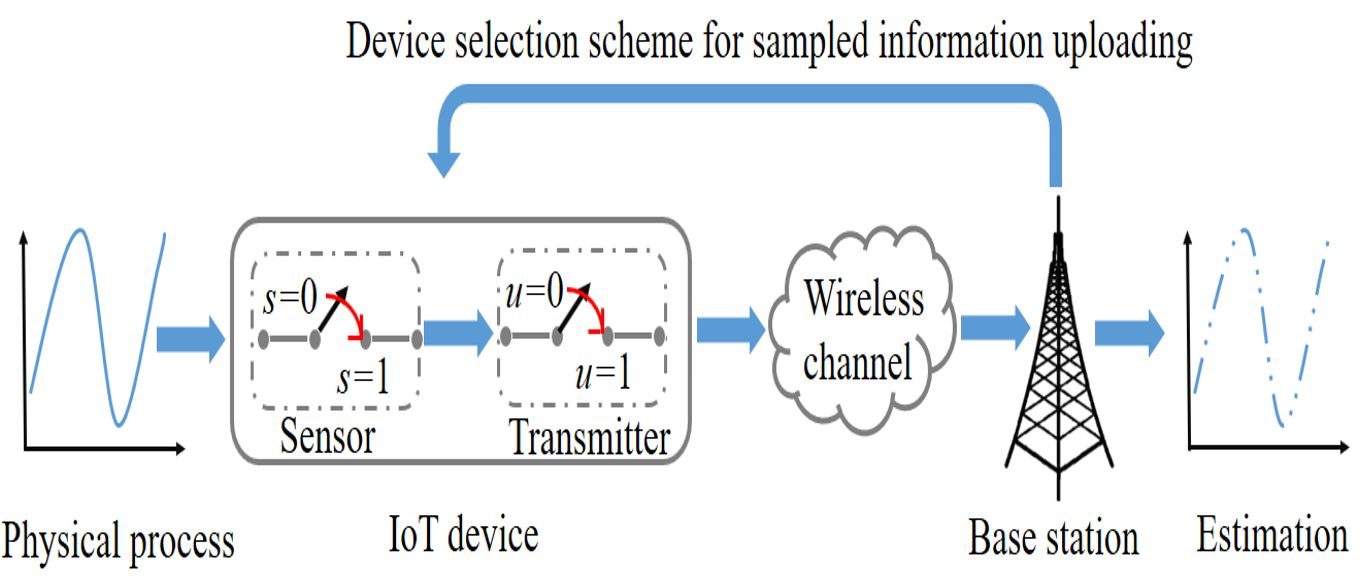}
\centering
\caption{An illustration of the considered IoT network.}
\label{fig2}
\end{figure}
Consider a real-time IoT system that consists of a BS and a set $\mathcal{M}$ of $M$ distributed IoT devices. In the considered model, each IoT device is equipped with a sensor and a transmitter. In particular, the sensor is used to monitor the real-time status of a physical system (e.g., an atmospheric sampler that monitors the variation of the atmospheric environment) and the transmitter is used to send the monitored information to the BS through a wireless channel, as illustrated in Fig. 1. Next, we first introduce the model of the physical process. Then, we introduce the AoI model to measure the freshness of monitored information of the physical process at IoT devices and the BS, respectively. Table I provides a summary of the notations used
throughout this paper.


\begin{table}[t]
\centering
\renewcommand\arraystretch{1}
\small
\caption{\normalsize Notation}
	\begin{tabular}{ | c | c ||}
    \hline  Notation & Description  \\
	\hline  $M$ & Number of devices  \\
	\hline  $\bm x_{m,t}$ & Dynamics of the physical process \\
    \hline  $\hat{\bm x}_{m,t}$ & Estimation of the physical process  \\
    \hline  $\bm A_m$ & Linear coefficient matrix  \\
    \hline  $\bm \epsilon_{m,t}$ & Random process \\
    \hline  $\bm y_{m,t}$ & Estimation error \\
    \hline  $\Omega_{m,t}$ &  Maximum variation frequency of the dynamics \\
    \hline  $\Delta_{m,t}$ & Maximum sampling interval \\
    \hline  $\xi_m$ &  Minimum sample frequency \\
    \hline  $\bm s_t$ & Sampling action vector \\
    \hline  $\bm u_t$ & Resource allocation vector \\
    \hline $\tau$ & Duration of each time slot $m$ \\
    \hline $P_{T}$ & Transmission power of each device  \\
    \hline  $Z_m$ &  Data size of each sampled packet \\
    \hline  $I$ & Number of resource blocks \\
    \hline  $l_{m,t}$ & Uplink transmission delay \\
    \hline  $\phi_{m,t}$ & AoI at device $m$  \\
    \hline  $ \Phi_{m,t}$ & AoI at the BS \\
    \hline  $C_s$ & Sampling cost for each packet \\
\hline $e_m$ &  Energy consumption  \\
\hline  $\gamma_E$ & Weighting parameter of energy \\
\hline  $\gamma_A$ & Weighting parameter of AoI \\
\hline
	\end{tabular}
\vspace{-0.6cm}
\end{table}




\subsection{Model of Physical Process}
We consider heterogeneous nonlinear time-varying dynamics to describe the variation of the physical process monitored by the IoT devices. These dynamics of the physical process over discrete time $t$ can be given by \cite{PHY}
\begin{equation}\label{eq:phy}
\begin{aligned}
{\bm x}_{m,t+1}={\bm A_m} \bm x_{m,t} + f_m(\bm x_{m,t}) + \bm \epsilon_{m,t},
\end{aligned}
\end{equation}
where $\bm x_{m,t}\in \mathbb{R}^{Z_m}$ is the system state vector sampled by device $m$ at time slot $t$ with $Z_m$ representing the data size of status information of device $m$ and $\bm \epsilon_{m,t}$ is an bounded disturbance independent of the system state. $f_m(\cdot):\mathbb{R}^{Z_m}\!\rightarrow\!\mathbb{R}^{Z_m}$ is a nonlinear function satisfying $f_m(\bm 0) = \bm 0$. $\bm A_m$ is a constant matric related to the linear dynamic systems. Note that, (1) has been widely used to model the physical process of nonlinear dynamic systems such as wide-area irrigation systems, electric power grids, automated highway systems, and environmental detection systems. For example, the dynamics of the atmospheric environment quality can be captured by (1) with $\bm x_{m,t}$ being the current air pollution index and $\bm x_{m,t+1}$ being the dynamics of the air pollution index while $\bm A_m \bm x_{m,t}$ and $f_m(\bm x_{m,t})$ represent the linear and nonlinear function to capture the effects of wind and precipitation. Using (\ref{eq:phy}), the current system state can be estimated based on the latest observed state, which is given by \cite{MQX}\vspace{-0.2cm}
\begin{equation}\label{eq:PHYest}
\begin{aligned}
\hat{\bm x}_{m,t}=\!{\bm A}_m^{\delta(t)}{\bm x}_{m,t-\delta(t)}\!+\!\sum\limits_{q = 1}^{\delta(t)}\!{\bm A}_m^{q-1}f_m(\bm x_{m,t-q}),
\end{aligned}
\end{equation}
where ${\bm x}_{m,t-\delta(t)}$ is the latest status information generated at time slot $t-\delta(t)$ with $\delta(t)$ being the duration of the generation time between ${\bm x}_{m,t}$ and ${\bm x}_{m,t-\delta(t)}$. Given the estimation of the system state vector at time slot $t$, the state estimation error can be expressed as\vspace{-0.2cm}
\begin{equation}\label{eq:PHYerror}
\begin{aligned}
\bm y_{m,t}=\hat{\bm x}_{m,t}\!-{\bm x}_{m,t}.
\end{aligned}
\end{equation}\vspace{-0.5cm}

In fact, $\bm y_{m,t}$ measures the estimation error of the current dynamics using the generated  physical process model. Using $\bm y_{m,t}$, each device can determine the sampling frequency at each time slot. To this end, we first need to calculate the maximum variation frequency of the physical process by analyzing the nonlinear dynamics of the physical system. For this purpose, (\ref{eq:PHYerror}) can be linearly approximated by \cite{Guo}
\begin{equation}\label{eq:PHYde}
\begin{aligned}
\frac{d\bm y_{m,t}}{dt}=\left(\bm A_{m,t}\!+\!{\bm J}_{f_m}({\bm x}_{m,t})\right)\cdot\bm y_{m,t}+\textsl{o}(\Vert\bm y_{m,t} \Vert),
\end{aligned}
\end{equation}
where $\left(\bm A_{m,t}\!+\!{\bm J}_{f_m}({\bm x}_{m,t})\right)\cdot\bm y_{m,t}$ is a first-order approximation with ${\bm J}_{f_m}({\bm x}_{m,t})$ being the Jacobian matrix of function $f_m$ and $\textsl{o}(\Vert\bm y_{m,t} \Vert)$ a high-order approximation that can be neglected compared to $\left(\bm A_{m,t}\!+\!{\bm J}_{f_m}({\bm x}_{m,t})\right)\cdot\bm y_{m,t}$. Then, we diagonalize $\bm A_{m,t}\!+\!{\bm J}_{f_m}({\bm x}_{m,t})$ to obtain the maximum variation frequency of the physical process at time slot $t$, which is given by
\begin{equation}\label{eq:PHYjaco}
\begin{aligned}
\bm A_{m,t}\!+\!{\bm J}_{f_m}({\bm x}_{m,t})={\bm U} \cdot \textrm{diag}(\mu_{1,t},\cdots,\mu_{Z_m,t}) \cdot \bm U^{-1},
\end{aligned}
\end{equation}
where $\textrm{diag}(\mu_{1,t},\!\cdots\!,\mu_{Z_m,t})$ is a diagonal matrix with $(\mu_{1,t},\!\cdots\!,\mu_{Z_m,t})$ being the eigenvalues and $\bm U=[\bm u_{1},\cdots,\bm u_{Z_m}]$ is a non-singular matrix with $\bm u_{z_m}\in\mathbb{R}^{Z_m}$ being the corresponding eigenvectors of $\bm A_{m,t}\!+\!\bm J_{f_m}(\bm x_{m,t})$. Based on (\ref{eq:PHYjaco}), the time-domain maximum variation frequency of the physical process can be computed as \cite{Guo}
\begin{equation}\label{eq:cutoff}
\begin{aligned}
\Omega_{m,t}\!=\!\!\!\!\!\!\!\mathop{\max}\limits_{z_{m,t}\in \mathcal{Z}_{m,t}}\!\!\!\!\!\!\!\left| {\rm Im}[\mu_{z_{m,t}}] \right|
\!\!+\!\!\sqrt{\!\frac{\Vert {\bm y}_{m,t}\Vert^{2}_{ 2}\!\!+\!\!\Vert \bm \epsilon_{m,t}\Vert^{2}_{ 2}}{\xi_m^2}\!-\!\!\!\!\!\!\!\mathop{\min}\limits_{z_{m,t}\in \mathcal{Z}_{m,t}}\!\!\!\!\!\!{\rm Re}[\mu_{z_{m,t}}]^2}\!,
\end{aligned}
\end{equation}
where $\mathcal{Z}_{m,t}$ is the set of $z_{m,t}$ and $\xi_m$ is a minimum frequency that device $m$ can distinguish, ${\rm Im}[\mu_{z_{m,t}}]$ and ${\rm Re}[\mu_{z_{m,t}}]$ are the imaginary part and real part of $\mu_{z_{m,t}}$, respectively. By assigning the sampling frequency $F_{m,t}=\Omega_{m,t}/\pi$ based on Nyquist theory, the maximum sampling interval of the dynamic physical process $\Delta_{m,t}$ can be given by
\begin{equation}\label{eq:AOIinterval}
\begin{aligned}
\Delta_{m,t}= 1/F_{m,t}=\pi/\Omega_{m,t}.
\end{aligned}
\vspace{-0.1cm}
\end{equation}
From (\ref{eq:cutoff}) and (\ref{eq:AOIinterval}), we can see that the maximum sampling interval $\Delta_{m,t}$ is related to $\bm y_{m,t}$. As $\bm y_{m,t}$ increases, the maximum variation frequency $\Omega_{m,t}$ increases and hence, $\Delta_{m,t}$ decreases. This is due to the fact
that, as the state estimation error $\bm y_{m,t}$ increases, (\ref{eq:phy}) cannot describe the physical process accurately, which imples device $m$ must increase its sampling frequency so as to collect more status information to capture the variation in the physical process and correct (\ref{eq:phy}).

\subsection{AoI Models for IoT Devices}
Different from the existing studies \cite{RDY}--[7] in which the AoI at edge device only depends on the time interval $\delta_m(t)$ between the consecutive  sampling status, in this paper, we consider the dynamic sampling frequency of a real-time physical system, thus, the AoI at each device $m$ will be affected by the maximum sampling interval $\Delta_{m,t}$ and $\delta_m(t)$, which is given by
\begin{equation}\label{eq:AOIdevice}
\vspace{-0.1cm}
\begin{aligned}
\!\!\phi_{m,t}(s_{m,t})
=\left\{ \begin{array}{l}
{\rm max}\{0,\delta_m(t)\!-\!\Delta_{m,t}\}, \;\;\; {\rm if}\; s_{m,t}=1,\\
{\rm min}\{\phi_{m,t-1}\!+\! \tau,\phi_{\rm max}\},\;\;\rm{otherwise,}
\end{array} \right.\
\end{aligned}
\end{equation}
where ${\rm max}\{0,\delta_m(t)\!-\!\Delta_{m,t}\}$ represents the AoI for device $m$ after sampling the physical process (i.e., $s_{m,t}\!=\!1$) with $\delta_m(t)\!-\!\Delta_{m,t}$ being the time interval between the current sampling action and the maximum sampling interval of the physical process. Meanwhile, ${\rm min}\{\phi_{m,t-1}\!+\! \tau,\phi_{\rm max}\}$ represents the AoI for device $m$ that does not sample the physical process (i.e., $s_{m,t}\!=\!0$) with $\tau$ being the duration of each time slot and $\phi_{\rm max}$ being the maximum sampling interval. From (\ref{eq:AOIdevice}), we can see that, as device $m$ samples the physical process at time $t$ and the time interval $\delta_m(t)$ is smaller than the maximal sampling interval $\Delta_{m,t}$ (i.e., $\delta_m(t)\!-\!\Delta_{m,t}\!<\!0$), the AoI at device $m$ decreases to zero. This is because when $\delta_m(t)$ is smaller than $\Delta_{m,t}$, the sampling frequency of device $m$ satisfies the constraint of the Nyquist-Shannon sampling theorem and thus, the sampled information can accurately represent the variation of the dynamic physical process. In contrast, when $\delta_m(t)\!-\!\Delta_{m,t}\!>\!0$, the AoI at device $m$ decreases to $\delta_m(t)\!-\!\Delta_{m,t}$. This is due to the fact that the sampling frequency of device $m$ cannot satisfy the constraint of the Nyquist-Shannon sampling theorem.



\subsection{AoI Model for the BS}
After generating the sampling information at time slot $t$, device $m$ requests to the BS for sampled information transmission. In the considered system, each device can only sample the monitored physical process after sending the current sampled information. An orthogonal frequency division multiple access (OFDMA) transmission scheme is used for sampled information transmission. We assume that the BS can allocate a set $\mathcal{I}$ of $I$ uplink orthogonal RBs to the devices. We also assume that each RB can be allocated to at most one device. The data rate of device $m$ transmitting sampled information to the  BS over each RB $i$ is
\begin{equation}\label{eq:uplinkdatarate}
r_{m,t}\left(u_{m,t}\right)=u_{m,t} W{\log _2}\left(\!1\!+\! {\frac{{{P_T}{ h_{m,t}}}}{{ \sigma^2_{\emph{N}} }}} \!\right)\!,
\end{equation}
where $W$ is the bandwidth of each RB, $P_T$ is the transmit power of each device $m$, and $u_{m,t}\in \{0,1\}$ is a device selection index at time $t$ with $u_{m,t}\!=\!1$ implying that device $m$ is selected by the BS to upload the sampled information at time slot $t$, and $u_{m,t}\!=\!0$, otherwise. $h_{m,t}$ is the channel gain between device $m$ and the BS. $\sigma^2_{\emph{N}}$ represents variance of the additive white Gaussian noise. Based on (\ref{eq:uplinkdatarate}), the uplink transmission delay between device $m$ and the BS is given by
\begin{equation}\label{eq:uplinkdelay}
l_{m,t}\left(u_{m,t}\right) =\frac{ Z_{m}}{r_{m,t}\left(u_{m,t} \right)}.
\end{equation}

Given the uplink transmission delay, the AoI at the BS for device $m$ can be expressed as
\begin{equation}\label{eq:AOIBS}
\begin{aligned}
\!\Phi_{m,t}(s_{m,t},u_{m,t})
\!=\!\left\{ \begin{array}{l}
\!\!\phi_{m,t}(s_{m,t})\!+\!l_{m,t}(u_{m,t}),\quad\;{\rm if} \; u_{m,t}=1,\\
\!\!{\rm min}\{\Phi_{m,t-1}\!+\! \tau,\Phi_{\rm max}\},\;\;\;\;\rm{otherwise,}
\end{array} \right.\
\end{aligned}
\end{equation}
where $\Phi_{\rm max}$ is the maximum sampling interval. From (\ref{eq:AOIBS}), we can see that, if device $m$ sends the sampled information to the BS at time slot $t$, then the AoI at BS will be updated to $\phi_{m,t}+l_{m,t}$, otherwise, the AoI increases by $\tau$.


Since the BS monitors multiple physical processes, we adopt the sum AoI at the BS as a scalar quantity to measure the information freshness. The randomness of each physical process will affect the AoI value and, hence, the average AoI is considered. We define the average sum AoI as
\begin{equation}\label{eq:AOISUM}
\begin{aligned}
\!\!\overline \Phi_{t}(\bm s_{t},\bm u_{t})
\!=\!\frac{1}{t}\sum\limits_{i = 1}^t \mathbb{E} \left[\sum\limits_{m = 1}^M \Phi_{m,i}(s_{m,i},u_{m,i})\right].
\end{aligned}
\end{equation}
where $\mathbb{E}[\cdot]$ is the expectation taken over the Rayleigh fading channel gain $g_{m,t}$, ${\bm s}_{t}=[s_{1,t},\ldots,s_{M,t}]$ and ${\bm u}_{t}=[u_{1,t},\ldots,u_{M,t}]$ are the sampling and the device selection vectors, respectively.

\subsection{Energy Consumption Model}



In our model, the energy used for each IoT device to sample and transmit the sampled information is
\begin{equation}\label{eq:energy}
\begin{aligned}
e_{m,t}(s_{m,t},u_{m,t})=s_{m,t}C_S+P_{T}l_{m,t}(u_{m,t}),
\end{aligned}
\end{equation}
where $s_{m,t}C_S$ is the energy consumption for sampling with $C_S$ being the cost for sampling the physical process and $P_{T}l_{m,t}(u_{m,t})$ is the energy consumption for transmitting the sampled information. Moreover, since the BS is supplied by a continuous power source, we do not consider the energy consumption of the BS. The average sum energy consumption is given by
\begin{equation}\label{eq:energySUM}
\begin{aligned}
\!\!\overline e_{t}(\bm s_{t},\bm u_{t})
\!=\!\frac{1}{t}\sum\limits_{i = 1}^t \mathbb{E} \left[\sum\limits_{m = 1}^M e_{m,i}(s_{m,i},u_{m,i})\right].
\end{aligned}
\end{equation}

\subsection{Problem Formulation}
Next, we introduce our optimization problem. Our goal is to minimize weighted sum of the AoI and energy consumption of all devices via optimizing the sampling vector $\bm s_t$ and the device selection vector $\bm u_t$. The optimization problem is given by
\begin{equation}\label{eq:max}
\begin{split}
\mathop {\min }\limits_{{\bm s}_{t},{\bm u}_{t}} \left(\gamma_{\rm A}\overline\Phi_{t}(\bm s_{t},\bm u_{t})+\gamma_{\rm E}\overline e_{t}(\bm s_{t},\bm u_{t}) \right)
\end{split}
\end{equation}
\vspace{-0.4cm}
\begin{align}\label{c1}
\setlength{\abovedisplayskip}{-20 pt}
\setlength{\belowdisplayskip}{-20 pt}
&\;\;\rm{s.\;t.}\;\scalebox{1}{${ s_{m,t},u_{m,t}} \in \left\{ {0,1} \right\},\forall {m} \in \mathcal{M},\forall {t} \in \mathcal{T},$}\tag{\theequation a}\\
&~~~~~~\sum\limits_{m \in \mathcal{M}}\!\! u_{m,t}  \leqslant I,\;\forall m \in \mathcal{M},\forall {t} \in \mathcal{T},\tag{\theequation b}
\end{align}
where $\gamma_{\rm A}$ and $\gamma_{\rm E}$ are the scaling parameters. (\ref{eq:max}a) guarantees that each device can sample the physical process once and can only occupy at most one RB for sampled information transmission at each time slot. (\ref{eq:max}b) ensures that each uplink RB can be allocated to at most one device.

The problem in (\ref{eq:max}) is challenging to solve by conventional optimization algorithms due to the following reasons. First, to find the optimal the sampling and transmission schemes using the traditional optimization algorithms, the BS must collect the information related to the dynamics of each monitored physical process. However, the dynamics of the physical process monitored by each device is not available for the BS until the devices sample the physical process and upload the sampled information to the BS successfully. Thus, each device must determine the sampling and transmission schemes based on the current dynamics of the monitored physical process. However, each device can only observe and analyze the local environment to determine its own policy. With partial observation, each device cannot find the optimal sampling and transmission schemes using the traditional optimization methods which require the dynamics of all monitored physical processes. In consequence, we propose a distributed RL algorithm that enables each device to use its local observation to estimate the Q-value under global observation and thus, cooperatively optimize the sampling and transmission schemes to minimize the weighted sum of the AoI and energy consumption.

\vspace{-0.2cm}
\section{QMIX Method for Optimization of Sampling Policy}
In this section, a novel distributed RL approach for optimizing the sampling policy $\bm s_t$ in (\ref{eq:max}) is proposed. In particular, the components of the proposed RL method is firstly introduced. Then, the process of using the proposed RL method to find the global optimal sampling policy for each device is explained. Given the sampling policy of each device, problem (\ref{eq:max}) is simplified and directly solved by dynamic programming. Finally, we analyze the convergence and complexity of the proposed RL method.

\subsection{Components of Distributed RL Method}
The proposed distributed RL method consists of six components: a) agents, b) actions, c) states, d) reward, e) individual value function, and f) global value function, which are specified as follows:

\begin{itemize}
\item \textbf{Agents}: The agents that perform the proposed RL algorithm are the distributed IoT devices. In particular, at each slot, each IoT device must decide whether to sample the physical process based on the local observation.


\item \textbf{Actions}: An action of each device $m$ is $s_{m,t}$ that represents the sampling policy of each device at time slot $t$. Thus, the vector of all devices' actions at time slot $t$ is ${\bm s}_t=[s_{1,t},\ldots,s_{M,t}]$.

\item \textbf{States}: An environment state is defined as $\bm o_t\!=\![\bm o_{1,t},\ldots,\bm o_{M,t}]$ where $\bm o_{m,t}\!\!=\!\![\phi_{m,t},F_{m,t},\!s_{m,t-1},u_{m,t-1}]$

represents the local observation of device $m$ with $\phi_{m,t}$ being the current AoI of device $m$  and $F_{m,t}$ being the current sampling frequency of device $m$. Here, $s_{m,t-1}$ and $u_{m,t-1}$ are the recorded historical sampling and transmission policy at time slot $t\!-\!1$, respectively. In the considered model, each device $m$ that is used to monitor a physical process can only observe its local state $\bm o_{m,t}$.



\item \textbf{Reward}: The reward of any sampling action on each device captures the weighted sum of the AoI and energy consumption resulting from the generation of the sampled information. Thus, the reward of each device can only be obtained when the sampled information is received by the BS successfully. To this end, the BS must first determine the device selection scheme ${\bm u}_t$, which can be found by the following theorem: 
    \begin{theorem}\label{thm1}{\rm Given the global state $\bm o_t$ and the sampling policy $\bm s_t$, the optimal device selection scheme $\bm u_t$ is given by:}
    \end{theorem}
    \vspace{-0.3cm}
    \begin{equation}
    \begin{aligned}
    \!\!&u_{m,t}^*
    =\left\{ \begin{array}{l}
    \!\!1,\quad\;{\rm if} \; m \in \mathcal M_1\\
    \!\!0 ,\quad\;{\rm otherwise}.
    \end{array} \right.\
    \end{aligned}
    \end{equation}
    where $\mathcal M_1=\{m\in\mathcal M|C_{m,t,1}<0\}$ with $C_{m,t,1}\!=\gamma_{\rm A}$\\$\!\left(\mathbb{E}\left[l_{m,t}(1)\right]\!+\!\phi_{m,t}(s_{m,t})\!-\!\Phi_{m,t-1}\!-\!\! \tau \right)\!+\!\gamma_{\rm E}P_{T}\mathbb{E}\left[l_{m,t}(1)\right]$.
    \vspace{0.1cm}
    \begin{IEEEproof}See Appendix A.
    \end{IEEEproof}\vspace{0.1cm}
    \quad From Theorem 1, we see that, the BS must first collect the global environment state $\bm o_t$ and the sampling actions of all devices $\bm s_t$ to find $\bm u_t$. Naturally, in the considered model, $\bm o_t$ and $\bm s_t$ can be obtained by the BS. This is because, as device $m$ samples the monitored physical process, the BS will receive a transmission request from device $m$ and allocate an RB to device $m$. Otherwise, the BS cannot receive the request message from device $m$. Based on the information of the received requests, the BS is aware of the sampling action of each device so as to obtain $\bm s_{t}$. Moreover, using the recorded information, the historical sampling action $s_{m,t-1}$ and device selection action $u_{m,t-1}$ are both available for the BS. In addition, after receiving the sampled information, the AoI of each device $\phi_{m,t}$ and the sampling frequency $F_{m,t}$ can be calculated using $s_{m,t}$, $s_{m,t-1}$, and $u_{m,t-1}$. Thus, the BS can obtain the global environment state $\bm o_t$. Given the global state $\bm o_t$ and the sampling policy $\bm s_t$, the optimal device selection scheme $\bm u_t$ can be determined. Based on Theorem 1, the reward function of each device is given by\vspace{-0.2cm}
    \begin{equation}\label{eq:reward}
    \begin{aligned}
    &\!\!R_{m,t}(\bm o_{m,t},\!{s}_{m,t})\!\!=\!\!-\!\left(\!\gamma_{\rm A}\!\Phi_{m,t}\!\left(\!s_{m,t},\!u_{m,t}\!\right)\!\!+\!\!\gamma_{\rm E}e_{m,t}\!\left(s_{m,t},\!u_{m,t}\!\right)\!\right)\!,
    \end{aligned}
    \end{equation}
    where $\gamma_{\rm A}\Phi_{m,t}(s_{m,t},u_{m,t})+\gamma_{\rm E}e_{m,t}(s_{m,t},u_{m,t})$ is the objective function of each device in (\ref{eq:max}) at each time slot. Note that, $R_{m,t}(\bm o_{m,t},{s}_{m,t})$ increases as the weighted sum of the AoI and energy consumption of device $m$ decreases, which implies that maximizing the reward of each device can minimize the weighted sum of the AoI and energy consumption.




\item \textbf{Individual value function}: The individual value function of each device $m$ is defined as $Q_m(\bm o_{m,t}, s_{m,t})$ which records the current local state and action and will be transmitted to the BS for the estimation of global value function. Based on the historical information recorded, each device can select the optimal sampling action $s_{m,t}$ using its local observation $\bm o_{m,t}$. However, due to the extremely high dimension of the state space with continuous variable $F_{m,t}$, it is computationally infeasible to obtain the optimal actions using the standard finite-state Q-learning algorithm \cite{Wang}. Hence, we adopt a DQN approach to approximate the action-value function using a deep neural network $Q_m(\bm o_{m,t},s_{m,t}|\bm \theta_m)$ where $\bm \theta_m$ is used to map the input local observation $\bm o_{m,t}$ to the output action $s_{m,t}$.



\item \textbf{Global value function}: We define a global value function $Q_{\rm tot}(\bm o_t, \bm a_t)$ that is generated by a mixing network $f(\cdot)$ in the BS to estimate all distributed devices' achievable future rewards at every global environment state $\bm o_t$. Different from \emph{value decomposition networks} (VDN) \cite{VDN} in which the global value function is defined as $\sum\limits_{m = 1}^M \!\! Q_m(\bm o_{m,t}, s_{m,t}|\bm \theta_m)$, we use a mixing network~to estimate the values of $Q_{\rm tot}(\bm o_t,\bm a_t)$ collected from distributed devices. The relationship between the global value function $Q_{\rm tot}(\bm o_t,\bm a_t)$ generated by the BS and $Q_m(\bm o_{m,t},s_{m,t}|\bm \theta_m)$ generated by each device can be given by
    \begin{equation}\label{eq:mix}
    \begin{aligned}
    &\!Q_{\rm tot}(\bm o_t,\bm a_t)\!\\
=&\!f\!\left(u_{1,t}Q_{1}(\bm o_{1,t}, s_{1,t}|\bm \theta_1),\ldots\!,u_{M,t}Q_{M,t}(\bm o_{M,t}, s_{M,t}|\bm \theta_M)\right)\\
=&\bm u_{t} \bm w_{t} \left[Q_1(\bm o_{1,t},\!s_{1,t}|\bm \theta_1\!),\!\ldots\!,\!Q_{\!M}\!(\bm o_{M,t},\!s_{M,t}|\bm \theta_M\!)\right] \!+\!\bm b_{t}\\
=&\!\!\sum\limits_{m = 1}^M\!u_{m,t}\!\left( w_{m,t} Q_{m,t}\left(\bm o_{m,t}, s_{m,t}|\bm \theta_m\right)\!+\!b_{m,t}\right)\\
    \end{aligned}
    \end{equation}
    where $f(\cdot)$ is the mixing network that is used to combine $Q_m(\bm o_{m,t}, s_{m,t}|\bm \theta_m)$ from each device $m$ monotonically with $\bm w_t\!=\![w_{1,t},\ldots\!,\!w_{M,t}]$ and $\bm b_t\!=\![b_{1,t},\ldots\!,\!b_{M,t}]$ being the weights and the biases of the mixing network, respectively. Here, we need to note that the value of $Q_m(\bm o_{m,t},s_{m,t}|\bm \theta_m)$ can only be obtained by the BS as device $m$ is selected. Otherwise, without an allocated RB, device $m$ cannot communicate with the BS and hence, the value of $Q_m(\bm o_{m,t}, s_{m,t}|\bm \theta_m)$ at device $m$ cannot be used to generate $Q_{\rm tot}(\bm o_t,\bm a_t)$ at the BS.



    \end{itemize}

\subsection{QMIX for Optimization of the Sampling Policy}
Given the components of the proposed QMIX algorithm, next, we introduce the entire procedure of training the proposed distributed QMIX algorithm to find the global optimal sampling policy and device selection schemes. The aim of the training process is to minimize the temporal difference (TD) error metric that is defined as follows
\begin{equation}\label{eq:QMIXloss}
\begin{aligned}
&\mathcal{L}(\bm \theta_1,\ldots,\bm \theta_M)\\
=&\mathbb{E}\! \left[ \!\left(\!Q_{\rm tot}(\bm o_t,\bm a_t)-R_t(\bm o_t,\bm a_t)-\gamma\mathop {\max }\limits_{{\bm a_t'}} Q_{\rm tot}(\bm o_{t+1}, \bm a_t) \!\right)^2 \!\right]\!,
\end{aligned}
\end{equation}
wher$R_t(\bm o_t,\bm a_t)\!=\!\!\sum\limits_{m = 1}^M \!\!u_{m,t} R_{m,t}(\bm o_{m,t},s_{m,t})$ and $\gamma$ is the discounted factor. To minimize the~TD error defined in (\ref{eq:QMIXloss}) in the distributed devices, we can observe the following: a) Given $Q_m(\bm o_{m,t}, s_{m,t}|\bm \theta_m)$ and $u_{m,t}$, calculating $Q_{\rm tot}(\bm o_t,\bm a_t)$ depends on the mixing network $f(\cdot)$ at the BS and b) Given $Q_{\rm tot}(\bm o_t,\bm a_t)$ and $\bm u_{t}$, updating $Q_{m,t}(\bm o_{m,t}, s_{m,t}|\bm \theta_m)$ only depends on the DQN network $\bm \theta_m$ at each device $m$. According to these observations, we can separate the training process of the proposed RL method into two stages: 1) BS training stage and 2) IoT device training stage, which can be given as follows:

\begin{figure}[t]
\centering
\setlength{\belowcaptionskip}{-0.45cm}
\vspace{-0.1cm}
\includegraphics[width=9cm]{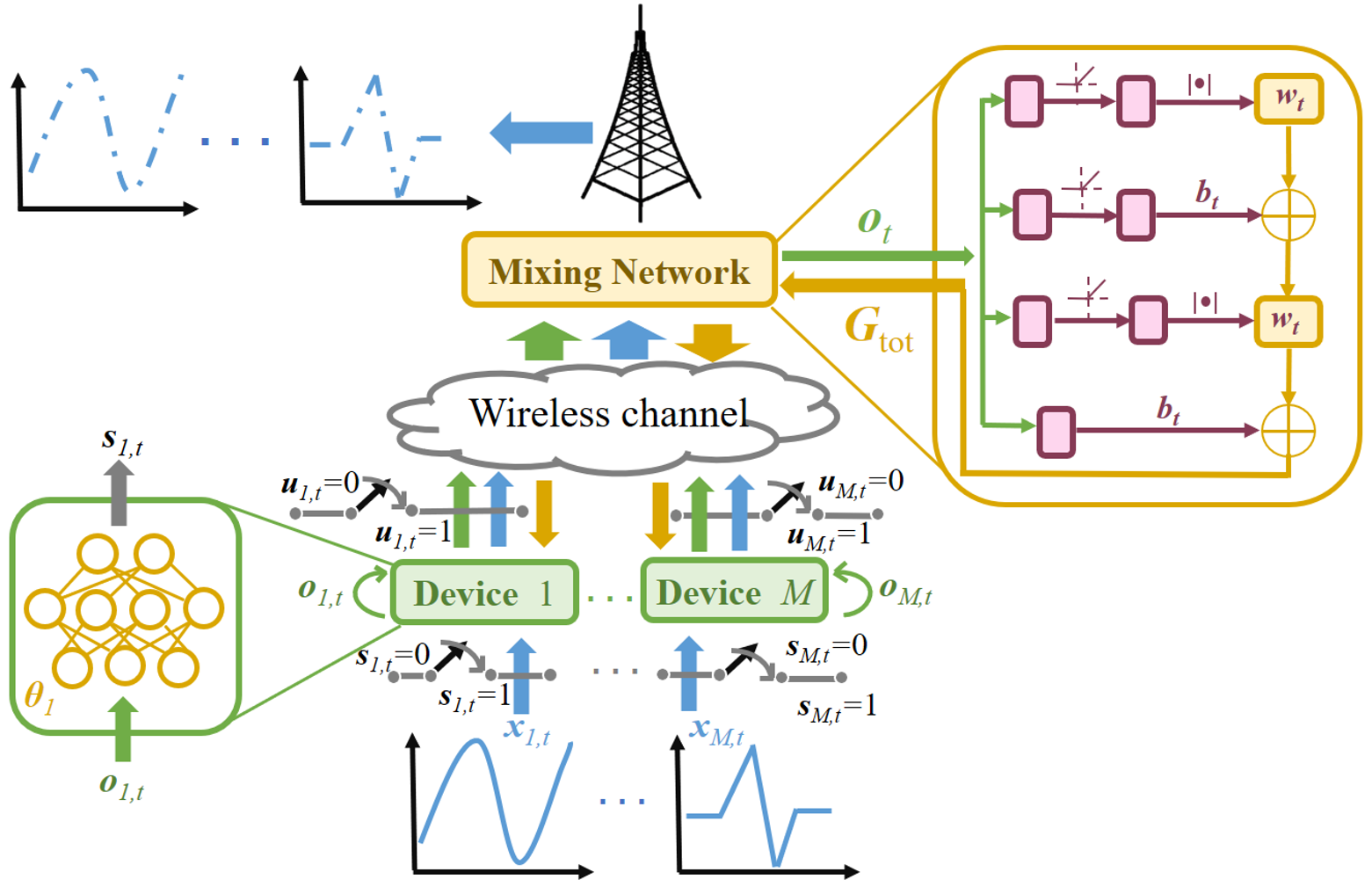}
\centering
\caption{The structure of the QMIX network.}
\vspace{-0.2cm}
\label{fig2}
\end{figure}

\begin{itemize}
\item \emph{BS training stage}: In this stage, the BS selects a subset of devices to transmit their sampled information and generate the global value function $Q_{\rm tot}(\bm o_t,\bm a_t)$. In particular, after executing the selected action $s_{m,t}$, device $m$ requests to the BS for sampled information transmission. Using the current and historical request information from all devices, the BS can obtain the global state $\bm o_t$ so as to determine the optimal device selection $\bm u_t$ based on Theorem 1. Given $\bm u_t$, the sampled information of the monitored physical process and the value of $Q_{m}(\bm o_{m,t}, s_{m,t}|\bm \theta_m)$ are collected by the BS. To estimate all devices' achievable future rewards, the BS must generate the global value function $Q_{\rm tot}(\bm o_t, \bm a_t)$ using the mixing network that takes $Q_{m}(\bm o_{m,t}, s_{m,t}|\bm \theta_m)$ and $\bm o_t$ as input. Given the weights and the biases at each linear layer of the mixing network, the BS can generate $Q_{\rm tot}(\bm o_t, \bm a_t)$ using (\ref{eq:mix}).





\item \emph{IoT device training stage}: In this stage, each IoT device must decide whether to sample the physical process. In particular, at each time slot, each device $m$ observes the local state $\bm o_{m,t}$ and chooses an action $s_{m,t}$. The probability of device $m$ selecting action $s_{m,t}$ is obtained via $\epsilon$ greedy method \cite{YFYRY}, which is given as follows: 
    \begin{equation}\label{eq:greedy}
    \begin{aligned}
    s_{m,t} \!=\!\! \left\{ \begin{array}{l}
    \!\!\!\! \mathop {\arg\!\max }\limits_{s_{m,t}\in {\mathcal{S}}}\!Q_m(\bm o_{m,t},\!s_{m,t}|\bm \theta_m),\; {\rm with \; probability}\;\epsilon,
    \\
    \quad \textrm{randint}(1,|\mathcal{S}|),{\kern 28pt}{\rm with \; probability}\;1\!-\!\epsilon,
    \end{array} \right.\
    \end{aligned}
    \end{equation}
    where $\epsilon$ is the probability of exploitation, $|\mathcal{S}|$ is the number of available actions for each device, and $\textrm{randint}(1,|\mathcal{S}|)$ is the random integer function that uniformly generates an integer ranging from 1 to $S$. As the monitored physical process is sampled by each device, a transmission request will be sent and thus, the BS can allocate the limited RBs for sampled information transmission and collect the local observation to generate the global value function $Q_{\rm tot}(\bm o_t, \bm a_t)$. As $Q_{\rm tot}(\bm o_t, \bm a_t)$ is received as feedback from the BS, an \emph{experience} defined in set $\mathcal{G}_m\!=\!\{\!(\bm o_{m,\!1}\!,\!\bm a_{m,\!1}\!,\!R(\!\bm o_{m,\!1},\!\bm a_{m,1})\!),\!...,\!(\bm o_{m,G}\!,\!\bm a_{m,G}\!,\!R(\!\bm o_{m,G}\!,\!\bm a_{m,G})\!)\!\}$ will be recorded by each device $m \in \mathcal{M}$. Then, each device selects a random batch $g_m$ from $\mathcal{G}_m$ to update its value function so as to accurately estimate future rewards. The update rule of the individual value function in each device can be given by
    \begin{equation}\label{eq:theta}
    \begin{aligned}
    &\Delta \bm \theta_m \!\\
=&\bm \theta_m^{i+1}\!-\!\bm \theta_m^{i}\\
    \!\!=&\alpha_m\nabla_{\! \bm \theta_m}\mathcal{L}(\bm \theta_1,\ldots,\bm \theta_M)\\
    \!\!=&\alpha_m \! \nabla_{\! \bm \theta_m}\!\!\left[\!\!\left(\!\!Q_{\rm tot}(\bm o_t,\!\bm a_t)\!-\!\!R(\bm o_t,\!\bm a_t)\!\!-\!\!\gamma\mathop {\max }\limits_{{\bm a_t'}}\! Q_{\rm tot}(\bm o_{t+1},\!\bm a'_t)\!\! \right)^{\!\!2}\right]\\
   \!\!= &\!2\kappa_{m,t}\Delta Q_{\rm tot}^2\!\nabla_{\! \bm \theta_m}\!Q_{m}(\bm o_{m,t},\! s_{m,t}|\bm \theta_m),
    \end{aligned}
    \end{equation}
    where $\Delta Q_{\rm tot}\!\!=\!\!Q_{\rm tot}\!(\bm o_t,\!\bm a_t\!)\!-\!\!R(\bm o_t,\!\bm a_t)\!-\!\gamma\!\mathop {\max }\limits_{{\bm a_t'}} \!Q_{\rm tot}(\bm o_{t\!+\!1},\!\bm a'_t)$

and $\kappa_{m,t}\!\!=\!\!\alpha_m u_{m,t} w_{m,t}$ with $\alpha_m$ being the update~step size of $Q_m(\bm o_{m,t},\!s_{m,t}|\bm \theta_m)$ in each distributed device $m$. From (\ref{eq:theta}), we can see that, as device $m$ is allocated an RB, the BS can transmit the value of $Q_{\rm tot}(\bm o_t,\bm a_t)$ and $w_m$ that are generated by the mixing network using $\bm o_t$ and $Q_m(\bm o_m, s_m|\bm \theta_m)$ to the selected devices, thus, device $m$ can update its own sampling policy by choosing greedy actions with respect to $Q_m(\bm o_m, s_m|\bm \theta_m)$. Otherwise, the device cannot communicate with the BS to obtain $Q_{\rm tot}(\bm o_t,\bm a_t)$ and participate in the update with a global state information.

\end{itemize}

The entire process of training the proposed QMIX algorithm is shown in Algorithm 1. At the beginning of the algorithm, each device selects the sampling action using the initial individual value function $Q_{m}(\bm o_{m,t},\!s_{m,t}|\bm \theta_m)$. After that, the set of devices that sample the monitored physical processes request RB allocation from the BS. Using the transmission request, the BS can obtain the local observation $\bm o_{m,t}$ from each device $m$ to determine the optimal $\bm u_t$. Then, the BS can collect the sampled information and the value of $Q_m(\bm o_{m,t}, s_{m,t}|\bm \theta_m)$ from the selected devices in $\bm u_t$. The collected information of $Q_m(\bm o_{m,t}, s_{m,t}|\bm \theta_m)$ and $\bm o_{m,t}$ will be considered as an input for the mixing network in the BS that is used to calculate the global value function $Q_{\rm tot}(\bm o_t,\bm a_t)$. Given $Q_{\rm tot}(\bm o_t,\bm a_t)$ from the BS, the devices can update their value parameters $\bm \theta_m$ based on (\ref{eq:theta}) and determine the sampling action with the updated value functions so as to minimize the weighted sum of AoI and the energy consumption.


\begin{algorithm}[t]\small
\caption{QMIX method}
\label{table}
\begin{algorithmic} [1] 
\REQUIRE The environment state $\mathcal{O}$, the sampling action space $\mathcal{S}$.\\
\ENSURE The sampling and device selection policy.\\ 
\STATE Initialize $\bm \theta_m$ of each DQN in each IoT device $m$ and the weights of mixing network in the BS.
\FOR {iteration $i=1:H$}
\FOR {each device $m$}
\STATE Observe the current local observation $\bm o_{m,t}$.
\STATE Choose an action $s_{m,t}$ according to $Q(\bm o_{m,t},s_{m,t}|\bm \theta_m)$.
\ENDFOR
\STATE The BS observes the local state $\bm o_{m,t}$ from each device $m$ to obtain global state $\bm o_t=[\bm o_{1,t},\ldots,\bm o_{M,t}]$.
\STATE Given $\bm o_t$, the BS optimizes the device selection scheme $\bm u_t$ using Theorem 1 and collects the sampled information from each selected device $m$.
\STATE The BS generates the global value function $Q_{\rm tot}(\bm o_t,\bm a_t)$ and transmit it to the selected devices.
\FOR {each selected device $m$}
\STATE Receive $Q_{\rm tot}(\bm o_t,\bm a_t)$ and records the experience $(\bm o_{m,t},\bm a_{m,t},R(\bm o_{m,t},\!\bm a_{m,t}))$ in $\mathcal{G}_m$.
\STATE Choose a random batch $g$ from $\mathcal{G}_m$.
\STATE Update $\bm \theta_m$ using (\ref{eq:theta}).
\ENDFOR
\ENDFOR
\end{algorithmic}
\end{algorithm}

\subsection{Convergence and Complexity of the Proposed Algorithm}

\subsubsection{Convergence of the Proposed Algorithm}
In this section, we first prove that the proposed RL method converges. However, we cannot find the exact value that the proposed RL method reaches. Our goal is to show that the proposed algorithm will not diverge. For this purpose, we first introduce the following definition on the gap between the optimal global Q-function $Q^*(\bm o_t,\bm a_t)$ and the optimal QMIX $Q^*_{\rm tot}(\bm o_t,\bm a_t)$.
\begin{definition}
{\rm The gap between the optimal global Q-value and the optimal QMIX value is defined as}
\vspace{-0.2cm}
\begin{equation}\label{eq:QMIX1}
\begin{aligned}
\varepsilon(\bm o_t, \bm a_t)\!=\!Q^*(\bm o_t, \bm a_t)-Q^*_{\rm tot}(\bm o_t, \bm a_t),
\end{aligned}
\end{equation}
{\rm where} $Q^*(\bm o_t, \bm a_t)\!\!=\!\!\sum\limits_{\bm o_t'}\!\!P_{\bm a_t}(\bm o_t,\!\bm o_t')\!\!\left[\!R(\bm o_t,\bm a_t)\!\!+\!\gamma\!\mathop {\max }\limits_{\bm a_t' \in \mathcal{A}} \!Q^*(\bm o_t',\!\bm a_t')\!\right]$ {\rm with} $P_{\bm a_t}(\bm o_t,\bm o_t')$ {\rm being the transition probability matrix} {\rm and} \\ $Q^*_{\rm tot}(\bm o_t, \bm a_t)\!=\!\!f(u_{1,t}Q_1^*(\bm o_{1,t},\!s_{1,t}),\ldots,u_{M,t}Q_M^*(\bm o_{M,t}, s_{M,t}))$ {\rm with} $Q_m^*(\bm o_{m,t}, s_{m,t})$ {\rm being the optimal} $Q_m(\bm o_{m,t}, s_{m,t})$.
\end{definition}\vspace{-0.2cm}
From Definition 1, we can see that the relationship between the optimal value of Q-function and the QMIX approach can be captured by a constant $\varepsilon(\bm o_t, \bm a_t)$ at each $(\bm o_t, \bm a_t)$. Moreover, as verified in \cite{WMD}, the Bellman operator in traditional Q network is a contraction operator with respect to the sup-norm over the globally observable information $\mathcal{O} \times \mathcal{A}$. Next, we prove that the Bellman operator in QMIX method is also a contraction operator with respect
to the sup-norm over partially observable information $\{\mathcal{O_{\rm 1},\cdots,O_{\rm M}}\} \times \{\mathcal{A_{\rm 1},\cdots,A_{\rm M}}\}$, which implies the gap $\varepsilon(\bm o_t, \bm a_t)$ will not effect the contraction of the Bellman operator in QMIX method. 
\begin{lemma}
{\rm When $I\geqslant M$, the optimal QMIX function is a fixed point of a contraction operator ${\rm H}Q_{\rm tot}$ in the sup-norm with modulus $\gamma$, i.e., $\Vert {\rm H}Q_{\rm tot}^1-{\rm H}Q_{\rm tot}^2 \Vert_\infty \leqslant \gamma \Vert Q_{\rm tot}^1-Q_{\rm tot}^2 \Vert_\infty$.}
\end{lemma}
\begin{IEEEproof}See Appendix A.
\end{IEEEproof}\vspace{0.2cm}
\noindent From Lemma 1, we observe that the Bellman operator in QMIX method is $\gamma$-contractive in the sup-norm. Such a contraction property constructs a sequence of action-value functions $\{Q_{\rm tot}^i\}_{i\geqslant 0}$ where the initialization function $Q_{\rm tot}^0$ is arbitrary. Using Lemma 1, we next prove that the sequence $\{Q_{\rm tot}^i\}_{i\geqslant 0}$ will converge as the QMIX algorithm empirically learns from a batch of data iteratively.
\begin{theorem}\label{thm2}{\rm The QMIX will converge to $Q^*_{\rm tot}(\bm o_t, \bm a_t)$.}
\end{theorem}
\begin{IEEEproof}See Appendix B.
\end{IEEEproof}\vspace{0.2cm}
Theorem 2 shows that the QMIX learning method always converges to $Q^*_{\rm tot}(\bm o_t, \bm a_t)$. In addition, we can also see that the gap $\varepsilon(\bm o_t, \bm a_t)$ that depends on the initialized weights of the neural network affects the performance of the sampling policy, thus, the sampling policy $\bm s_t$ obtained by the QMIX algorithm may not be the optimal solution in (\ref{eq:max}). 

\subsubsection{Complexity of the Proposed Algorithm}
Next, we present the complexity for the optimization of device selection scheme and the training the QMIX algorithm that consists of distributed DQNs in each IoT device and a mixing network in the BS, which is detailed as follows.


\vspace{0.1cm}
a) First, we explain the complexity of optimizing the device selection, which lies in obtaining the optimal $\bm u_{t}^*$. According to Theorem 1, the complexity of optimizing device selection is $\mathcal O(MT)$.

\vspace{0.1cm}
b) In terms of complexity of training distributed DQNs, each device needs to update its own RL parameter $\bm \theta_m$ that is used to determine the sampling action $s_{m,t}$ based on the local observation $\bm o_m$. Hence, the time-complexity of training a DQN that is a fully connected network depends on the dimension of input $\bm o_{m,t}$ and output $s_{m,t}$, as well as the number of the neurons in each hidden layer \cite{WYN1}. Let $l_i$ denote the number of neurons of any given hidden layer $i$ and $L_{\rm D}$ denote the number of hidden layers in the DQN of device $m$, the time-complexity of the DQN is $\mathcal{O}(\sum\limits_{i = 1}^{L_{\rm D}} l_il_{i+1}+|\bm o_m|l_1+|s_m|L_{\rm D})$ where $|\bm o_m|$ and $|s_m|$ represent the dimension of $\bm o_{m,t}$ and $s_{m,t}$, respectively. 

\vspace{0.1cm}
c) In terms of complexity of training the mixing network, the BS first needs to generate the weights and the biases of the mixing network that is a feed-forward neural network. Hence, the time-complexity of the mixing network lies in the generation of the weights and the biases of each layer. In particular, the weights of each layer in the mixing network is generated by separate hypernetworks \cite{Hypernetworks} that consists of two fully-connected layers with a ReLU nonlinearity, followed by an absolute activation function, to ensure the non-negativity of the mixing network weights. Moreover, the biases are produced in the same manner but are not restricted to being non-negative. Hence, the time-complexity of generating the parameters in the mixing network is $\mathcal{O}(|\bm o_t|L_{\rm W}|\bm w_t| + L_{\rm B}|\bm b_t|)$ \cite{PYMY} where $L_{\rm W}$ and $L_{\rm B}$ denote the number of the neurons to product the weights and the biases, respectively. $|\bm w_t|$ and $|\bm b_t|$ represent the dimension of $\bm w_t$ and $\bm b_t$, respectively.

\section{Simulation Results and Analysis}\label{se:system}
In our simulations, we consider a circular network area having a radius $r = 100$ $m$. One BS is located at the center of the network area and $M = 20$ IoT devices are uniformly distributed. The data used to model real-time dynamics is obtained from the Center for Statistical Science at Peking University \cite{data}. Table II defines the values of other parameters. A uniform sampling policy and the traditional fully distributed DQN method are considered for comparison.

\begin{table}
\centering
\vspace{-0.3cm}
\renewcommand\arraystretch{1}
\caption{Simulation Parameters \cite{MZWCH}}
\vspace{-0.2cm}
\small
\setlength{\tabcolsep}{0.9mm}{
\begin{tabular}{|c|c|c|c|}
\hline
\textbf{Parameters}&\textbf{Values}&\textbf{Parameters}&\textbf{Values}\\
\hline
\emph{M}& 20 &$\tau$&1 s \\
\hline
\emph{I}& 10&$\sigma^2_N$& -95 dBm \\
\hline
\emph{W}& 180 kHz &$\xi_m$& 10 Hz \\
\hline
$P_T$& 0.5 W &$\gamma_{\rm A}$& 0.5 \\
\hline
$C_S$& 0.5 mJ& $\gamma_{\rm E}$& 0.5  \\
\hline
$\phi_{\rm max}$& 5 & $\Phi_{\rm max}$& 5 \\
\hline
$Z_m$&10 bit & $r$& 100\\
\hline
\end{tabular}}
\vspace{-0.3cm}
\end{table}

\begin{figure}[t]
\centering
\setlength{\belowcaptionskip}{-0.45cm}
\vspace{-0.1cm}
\includegraphics[width=8.3cm]{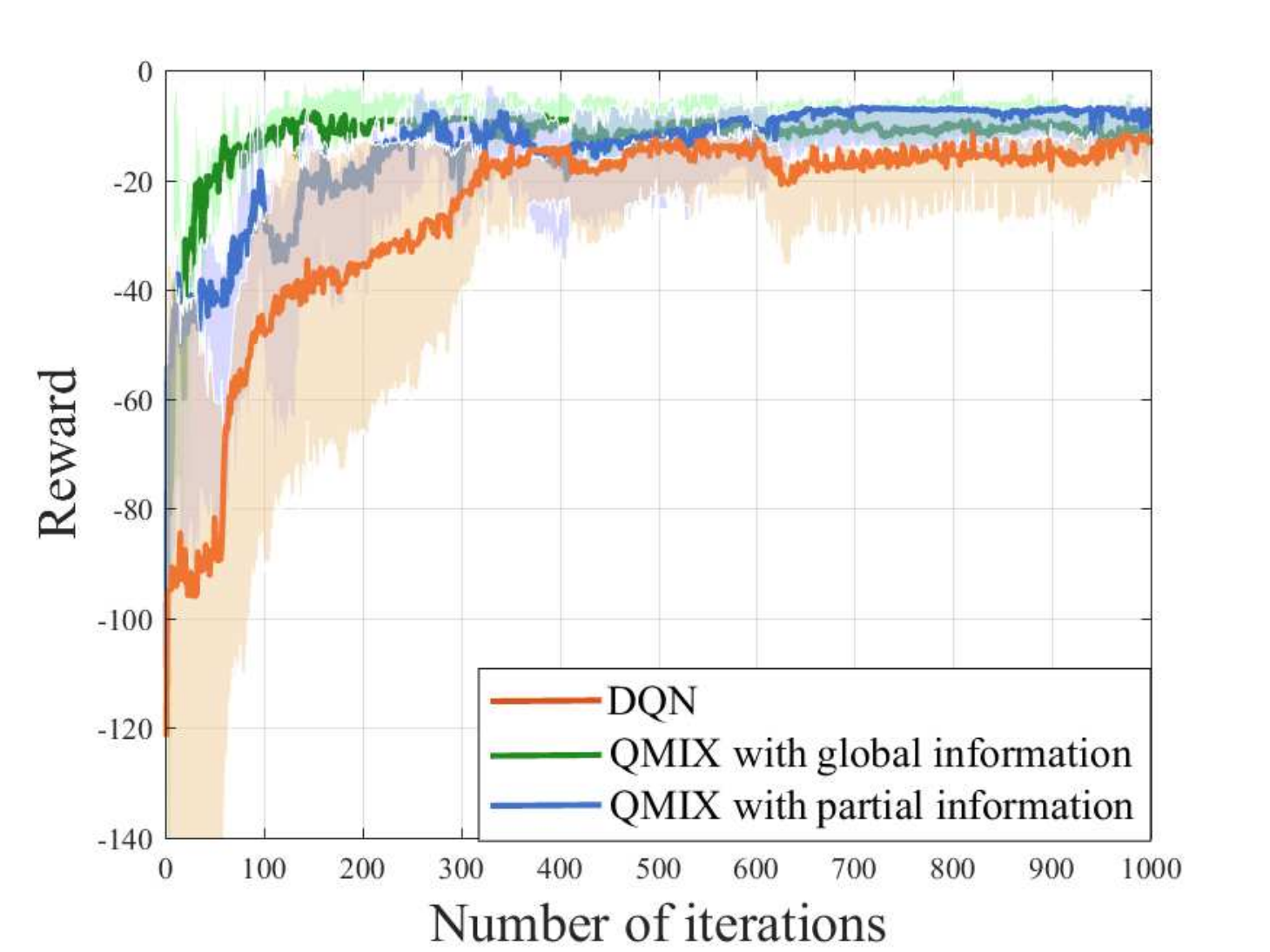}
\vspace{-0.2cm}
\centering
\caption{\small Value of the reward function as the total number of iterations varies.}
\vspace{-0.3cm}
\label{reward_aver}
\end{figure}

Fig. \ref{reward_aver} shows how the value of the reward function changes as the total number of iterations varies. In Fig. \ref{reward_aver}, the line and the shadow are the mean and the standard deviation computed for 20 users with 10 RBs. Due to the limited RBs, the BS cannot collect $Q_m(\bm o_m, s_m|\bm \theta_m)$ from all devices at each time slot. To investigate how the constraint on the number of RBs affects the performance of the QMIX algorithm, we compare the proposed QMIX with partial information of $Q_m(\bm o_m, s_m|\bm \theta_m)$ from selected devices to the QMIX with global information of $Q_m(\bm o_m, s_m|\bm \theta_m)$ from all devices in the same system. From Fig. \ref{reward_aver}, we can see that, compared to the traditional DQN algorithm, the proposed QMIX algorithm can achieve better performance at the beginning of the training process. This is because that at the beginning of the training process, the BS can collect the local observation at each device to generate the global value function that enables each device to quickly adjust the sampling policy based on the local state information, which results in a rapid improvement at the beginning of the training process. However, the proposed QMIX approach achieves a 38.1\% loss in terms of the number of iterations needed to converge compared to the QMIX algorithm with global information. This is due to the fact that, under a limited number of RBs, the BS can only collect a subset of $Q_m(\bm o_m, s_m|\bm \theta_m)$ from the selected devices to generate the global value function. With partial information, the proposed algorithm cannot capture the relationship between the sampling policies of all devices during one iteration, thus decreasing convergence speed. Fig. \ref{reward_aver} also shows that the proposed algorithm can achieve up to 24.4\% gains in terms of the weighted sum of AoI and energy consumption compared with the DQN algorithm. This implies that the proposed algorithm enables the devices to cooperatively train the learning models based on the estimation on the strategy outcomes, thus improving the performance of the sampling policies of the distributed devices.

\begin{figure}
\begin{minipage}{1\linewidth}
  \centerline{\includegraphics[width=8.2cm]{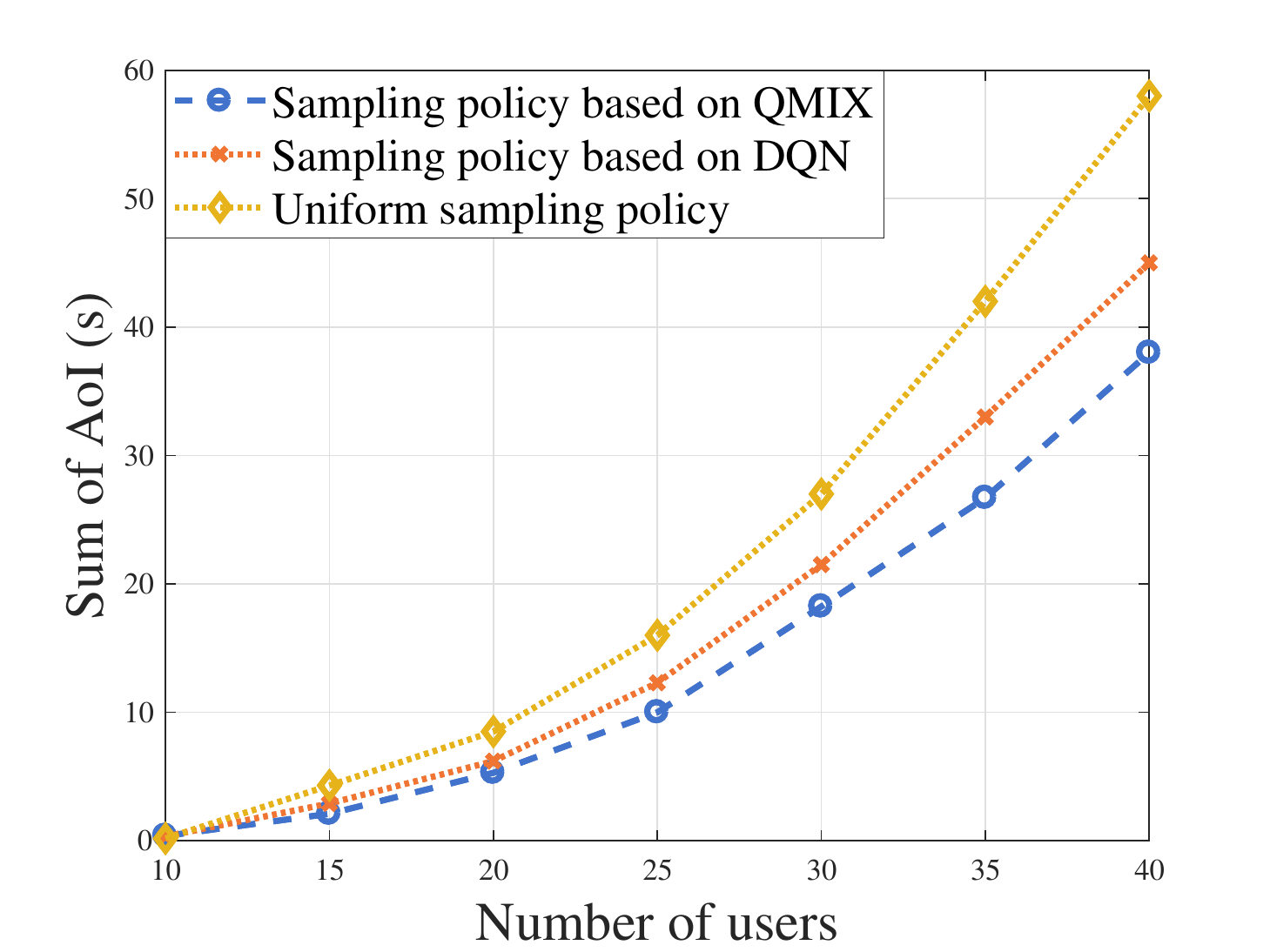}}
  \centerline{\footnotesize (a) The sum of AoI vs. the number of IoT devices.}
\end{minipage}
\qquad\quad\!\!\!\!
\begin{minipage}{1\linewidth}
  \centerline{\includegraphics[width=8.2cm]{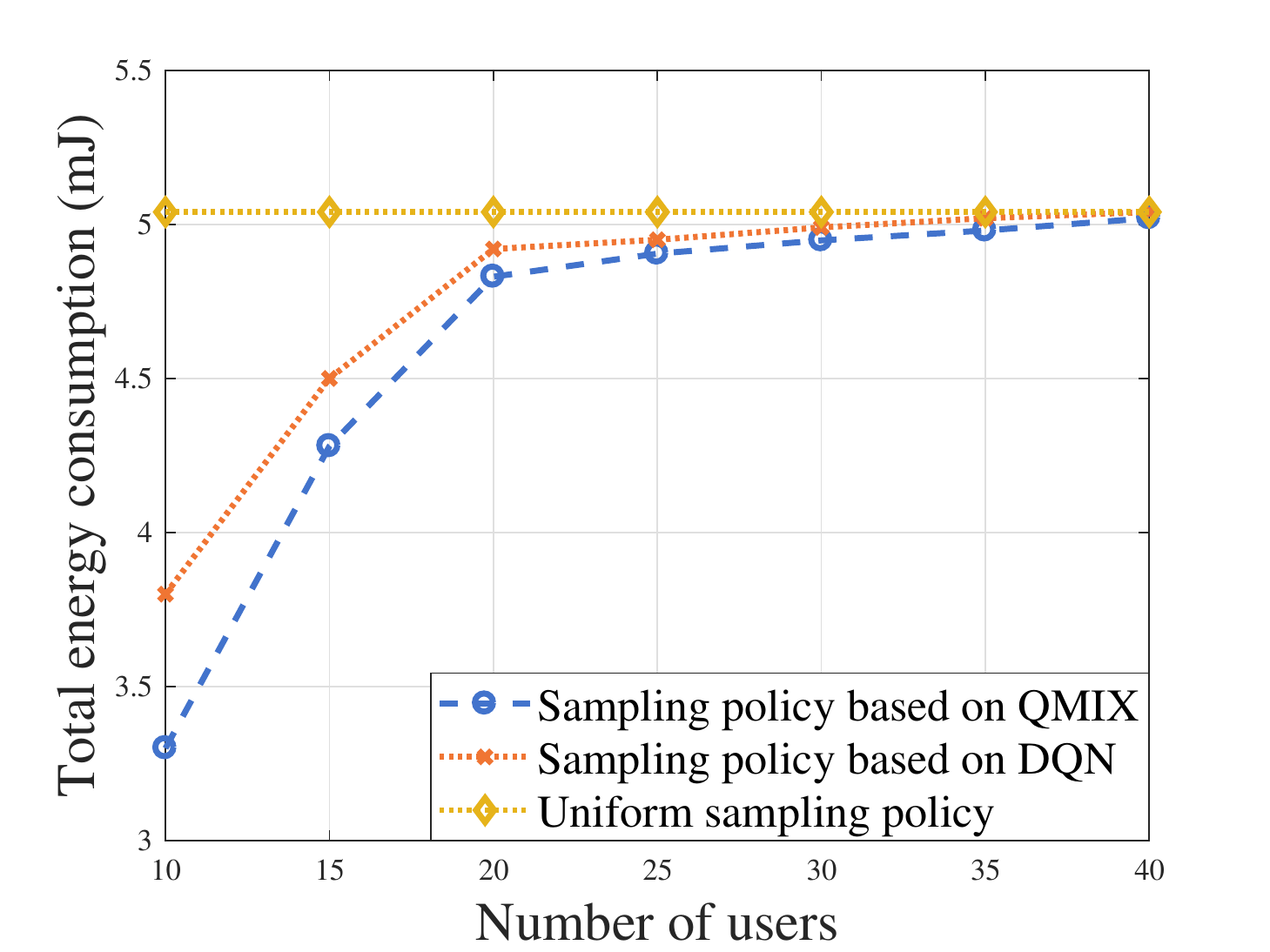}}
  \centerline{\footnotesize (b) The total energy consumption vs. the number of IoT devices.}
\end{minipage}
\caption{\small The sum of AoI and the total energy consumption as the number of IoT devices varies.}
\label{sum}
\vspace{-0.3cm}
\end{figure}

In Fig. \ref{sum}, we show how the sum of the AoI and the total energy consumption of all devices change as the number of edge devices varies. From Fig. \ref{sum}(a), we can see that, the sum AoI increases as the number of devices increases. This is due to the fact that, the number of RBs is limited in the considered system and, hence, as the number of devices increases, some devices may not be able to sample and transmit their monitored information to the BS immediately, thus resulting in an increase of the sum of AoI. Moveover, the sum AoI increases rapidly as the number of devices continues to increase. This is because, as the number of RBs is much smaller than the number of devices, most of the devices must wait until being allocated the RB so as to update the sampled information which results in a great growth in terms of AoI. In Fig. \ref{sum}(a), we can also see that the proposed algorithm can reduce the sum of the AoI by up to 17.8\% and 33.9\% compared to the sampling policy based on DQN and the uniform sampling policy, respectively, for the case with 10 RBs and 40 devices. This gain stems from the fact that the proposed algorithm enables the BS to observe the global state information so as to generate the global value function that enables each distributed device to achieve a better performance of the sample policy. From Fig. \ref{sum}(b), we can see that, as the number of devices increases, the total energy consumption increases. This is due to the fact that, as the number of devices increases, the number of devices that must sample the physical process and transmit the sampled information to the BS increases, and, hence, the total energy consumption for status sampling and uploading increases. Fig. \ref{sum}(b) also shows that, when the number of devices is larger than 20, the total energy consumption of all algorithms remains nearly constant because of the limited number of available RBs. From Fig. \ref{sum}(b), we can see that the proposed algorithm can reduce the total energy consumption by up to 13.2\% and 35.1\% compared to the sampling policy based on DQN and the uniform sampling policy for the case with 10 RBs and 10 devices. This gain stems from the fact that the proposed algorithm enables the BS to collect the information of the AoI and the physical dynamics from the distributed devices, thus being able to select less devices to reduce the sum of AoI and capture the variation of the physical process with less energy.

\begin{figure}
\begin{minipage}{1\linewidth}
  \centerline{\includegraphics[width=8cm]{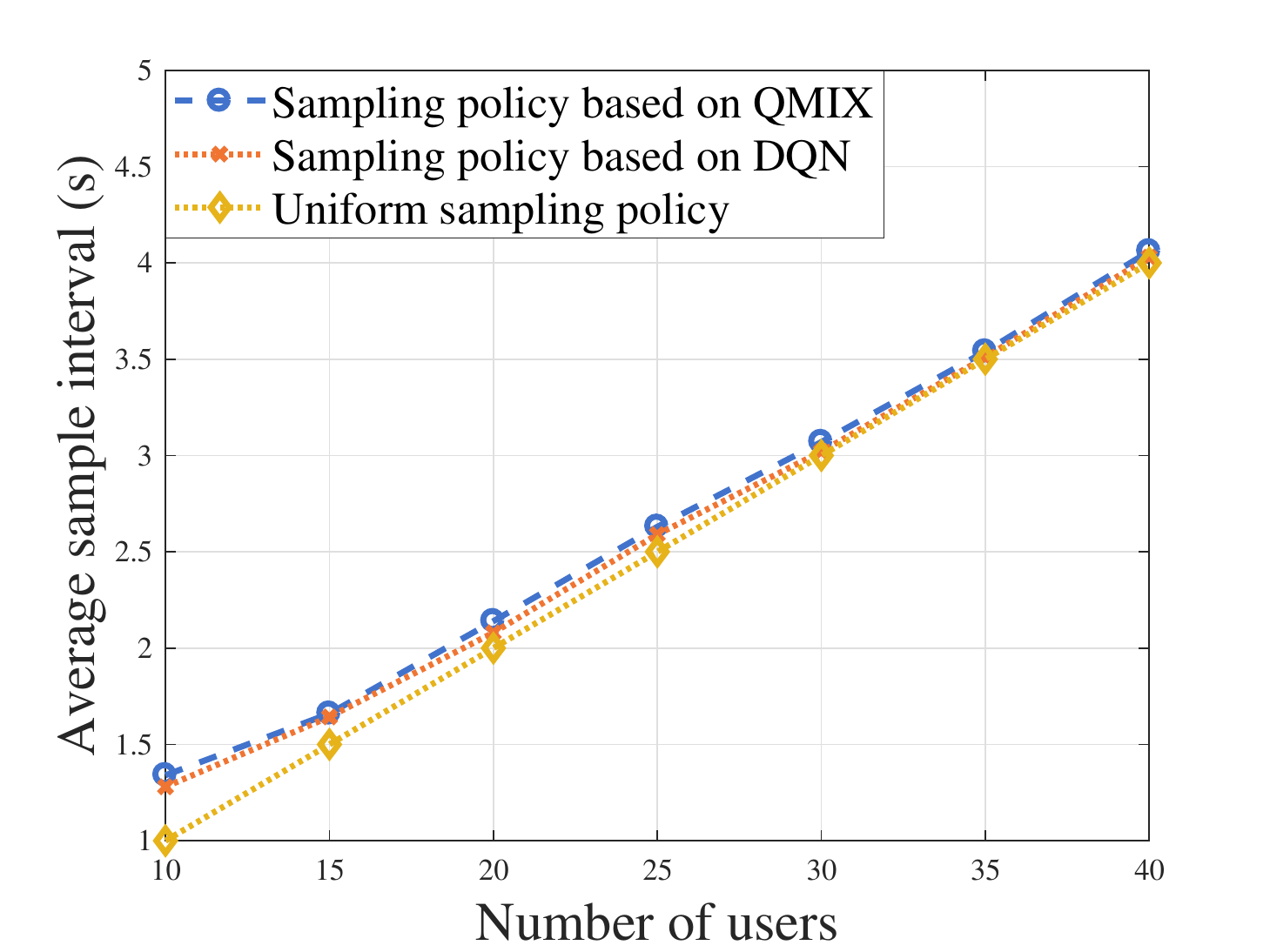}}
  \centerline{\footnotesize (a) The average sample interval vs. the number of IoT devices.}
\end{minipage}
\qquad\quad\!\!\!\!
\begin{minipage}{1\linewidth}
  \centerline{\includegraphics[width=8cm]{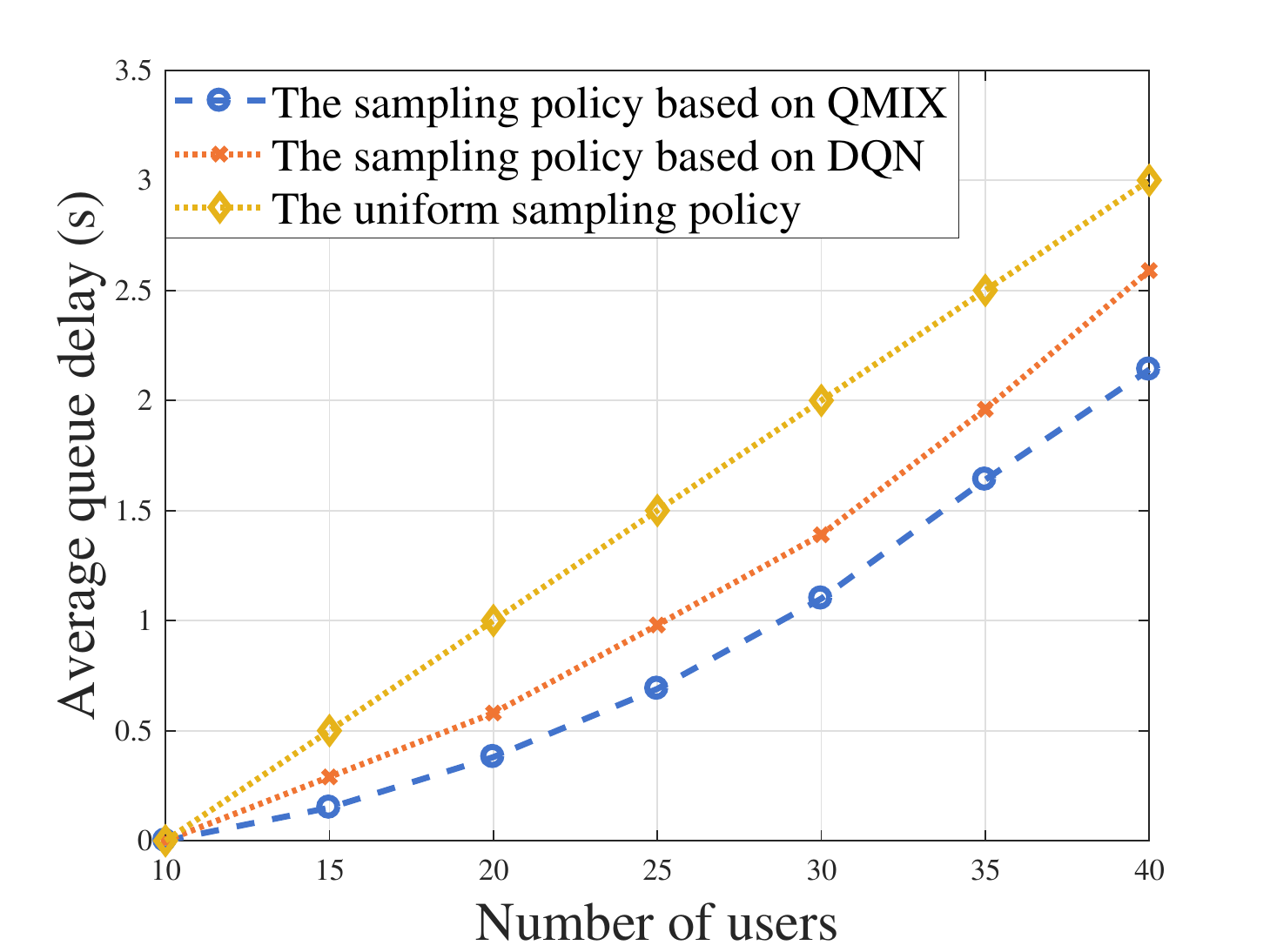}}
  \centerline{\footnotesize (b) The average queue delay vs. the number of IoT devices.}
\end{minipage}
\caption{\small The average sample interval and the average queue delay changes as the number of IoT devices varies.}
\label{interval}
\vspace{-0.4cm}
\end{figure}

Fig. \ref{interval} shows how the average sample interval and the average queue delay changes as the number of IoT devices varies. Clearly, from Fig. \ref{interval}(a), we can see that, as the number of devices increases, the average sample interval increases. This is because the number of RBs is limited and, hence, as the number of devices increases, the probability that each device can be allocated a transmission opportunity to upload the sampled status information decreases. In consequence, each device must increase the sample interval so as to decrease the number of sampled packets for energy saving. Fig. \ref{interval}(a) also shows that the average sample intervals of all algorithms are essentially identical. This is because that the devices are used to monitor different physical processes which changes as time elapses. For this purpose, all algorithms must fully utilize the limited energy and RBs for status sampling and uploading and hence, the average sample interval of all algorithms are basically same. From Fig. \ref{interval}(b), we can see that, as the number of users increases, the average queue delay of each sampled packet increases. This stems from the fact that, as the number of users increases, the BS can not be able to allocate limited RBs to all devices immediately, thus resulting in additional queue delay of each sampled packet. Fig. \ref{interval}(b) also shows that the proposed algorithm achieves up to 29.3\% and 15.9\% gains in terms of the average queue delay compared to the uniform sampling policy and the sampling policy based on DQN. This is because that the proposed algorithm enables the devices to learn the sampling policies from each other by using the global value function generated by the BS and hence, cooperatively sampling the physical process and reducing the queue delay.
\begin{figure}
\begin{minipage}{1\linewidth}
  \centerline{\includegraphics[width=7.75cm]{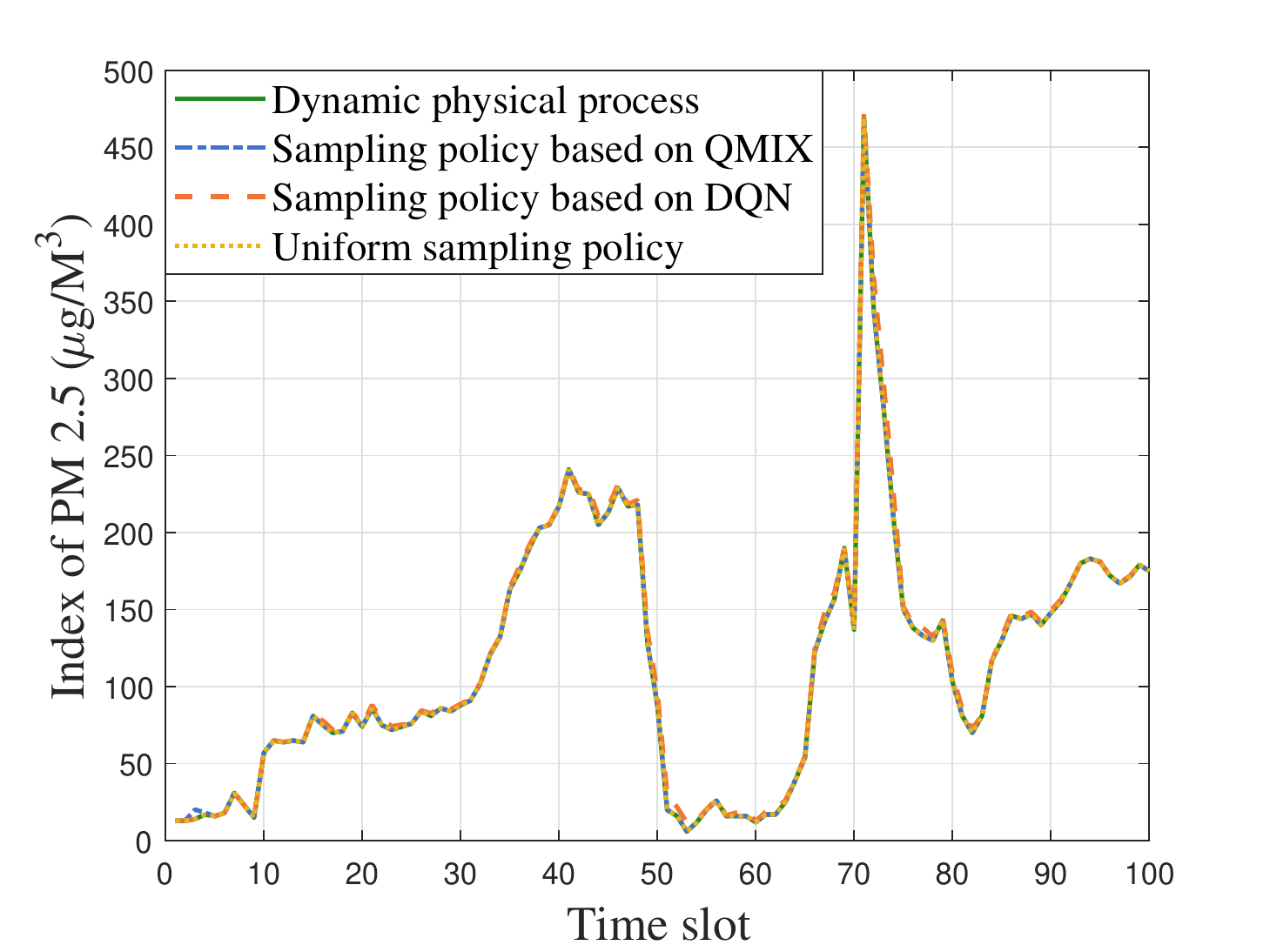}}
\vspace{-0.2cm}
  \centerline{(a) 10 users with 10 RBs}
  \vspace{-0.04cm}
\end{minipage}
\quad\quad\!\!\!\!\!
\vspace{-0.16cm}
\begin{minipage}{0.95\linewidth}
  \centerline{\includegraphics[width=7.25cm]{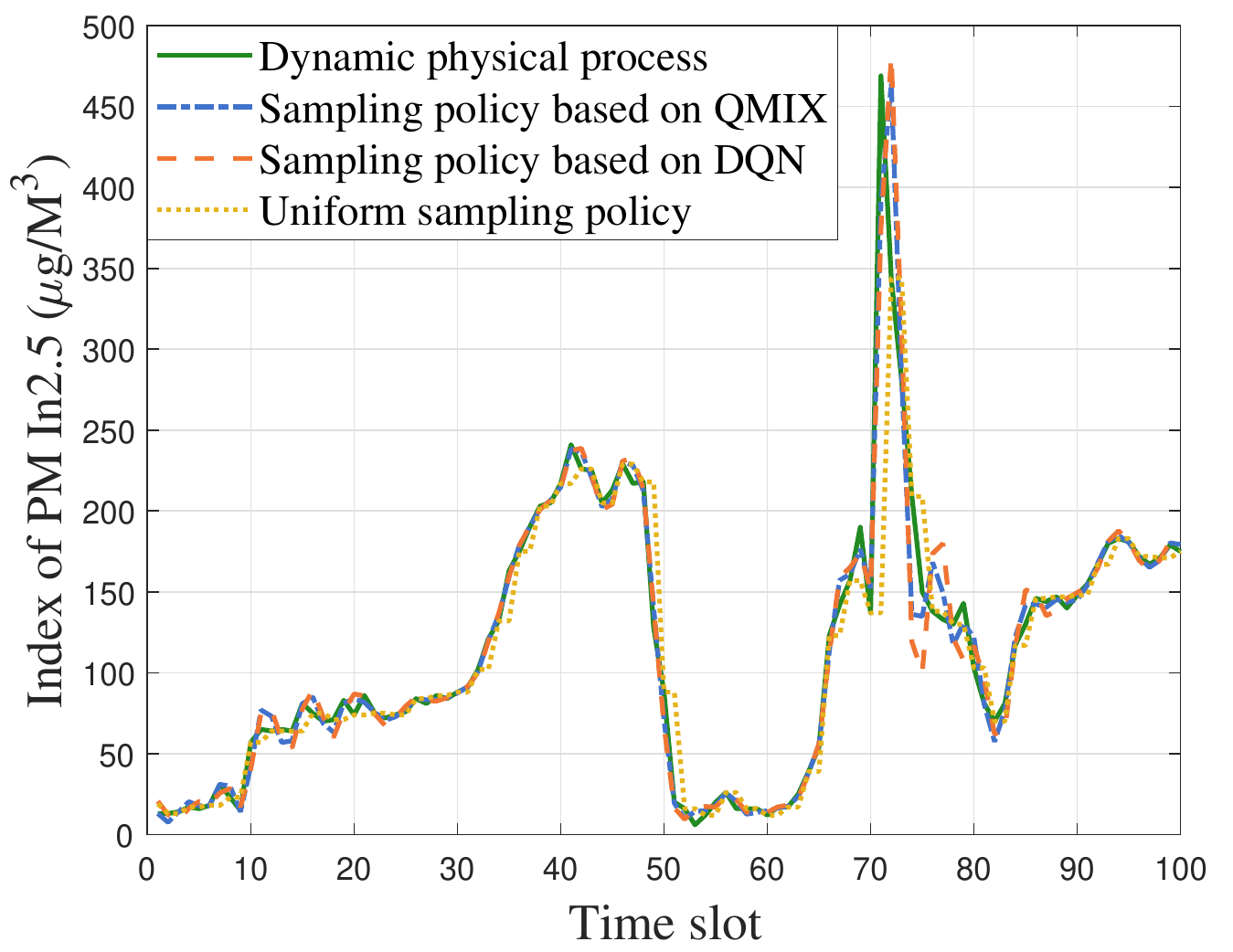}}
\vspace{-0.12cm}
  \centerline{(b) 20 users with 10 RBs}
\vspace{0.11cm}
\end{minipage}
\vfill
\begin{minipage}{0.95\linewidth}
  \centerline{\includegraphics[width=7.25cm]{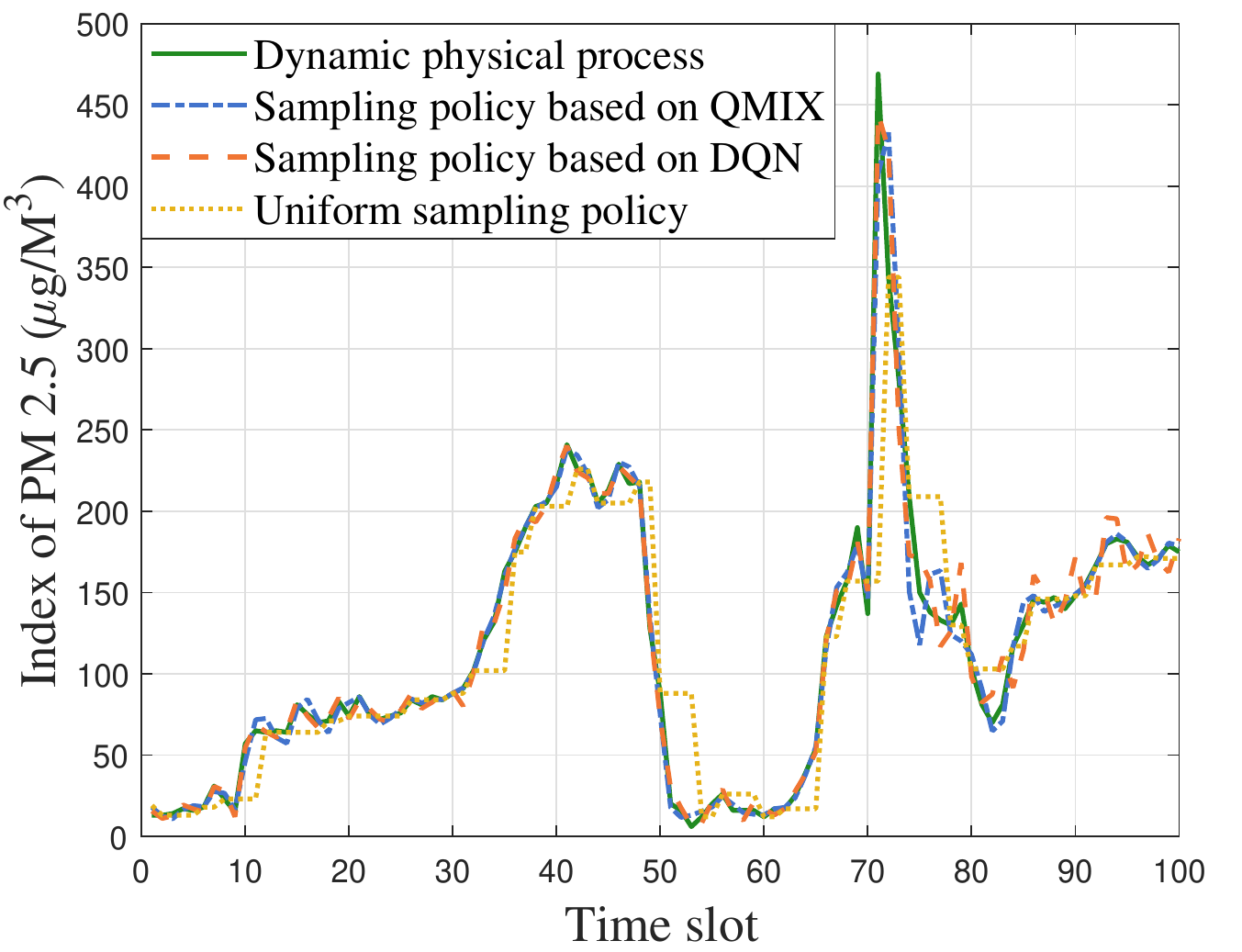}}
\vspace{-0.12cm}
  \centerline{(c) 30 users with 10 RBs}
\end{minipage}
\quad\quad\!\!\!\!\!
\begin{minipage}{0.95\linewidth}
  \centerline{\includegraphics[width=7.25cm]{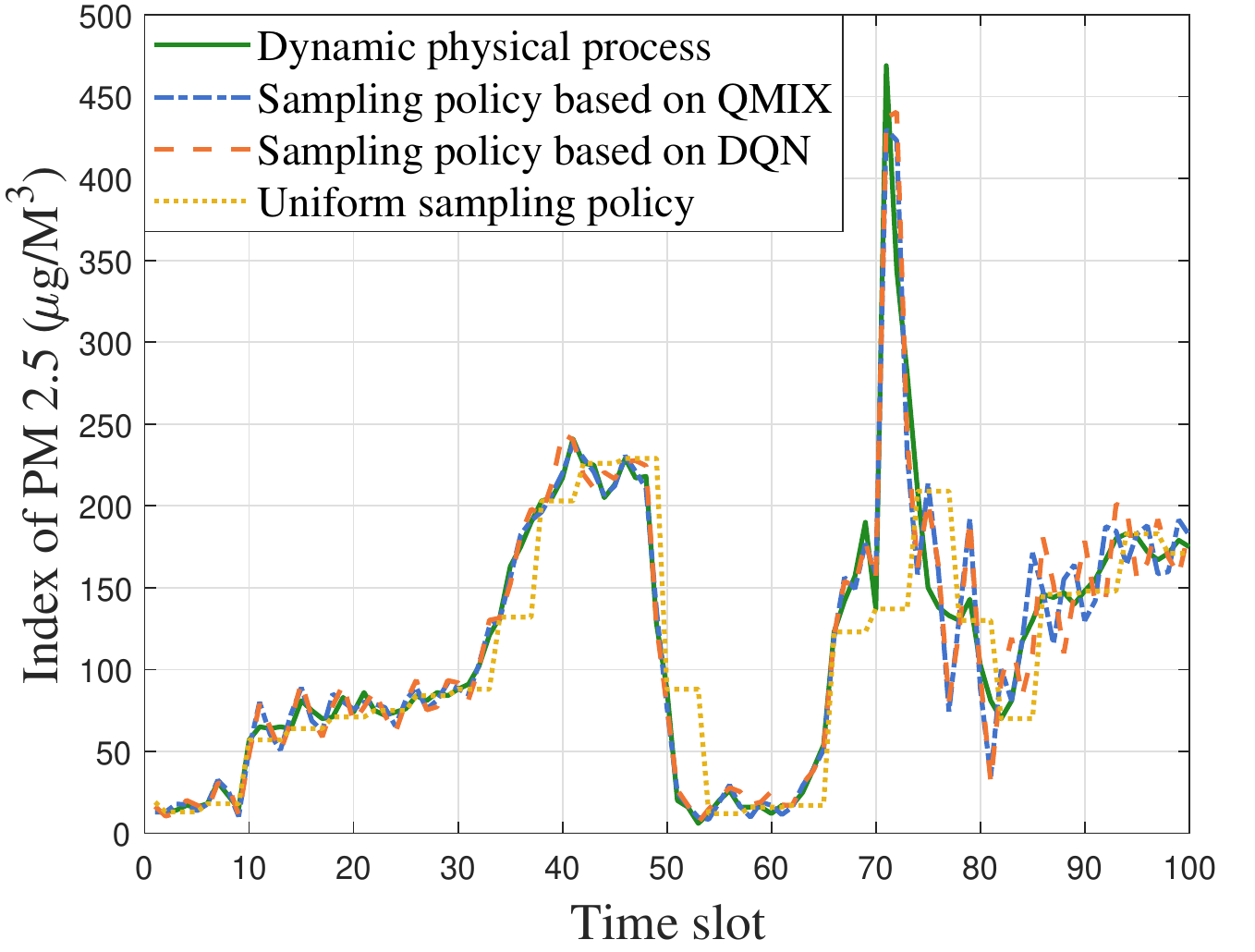}}
\vspace{-0.13cm}
  \centerline{(d) 40 users with 10 RBs}
\end{minipage}
\vspace{-0.13cm}
\caption{\small The estimation of the dynamics of the physical process based on different algorithms.}
\label{example}
\end{figure}

Fig. \ref{example} shows an example of the estimation of the physical process with different number of users. In this figure, we can see that, as the physical process varies rapidly, the estimation error of PM 2.5 increases. The reason is that, as the index of the PM 2.5 changes rapidly, each device must increase the sampling frequency so as to collect more status information to capture the variation of the physical process. However, with limited RBs, each device may not be able to transmit these sampled packet to the BS immediately and hence, the estimation error of the proposed approach increases. Fig. \ref{example} also shows that, as the index of the PM 2.5 changes slowly, the estimation error will not be reduced to zero, especially for the case with 40 devices. This stems from the fact that, as the number of devices is much lager than the number of RBs, the devices must cooperatively sample the different monitored physical process. Thus, as the index of the PM 2.5 changes slowly, each device will not sample this physical process so as to save the energy and the limited RBs that can be allocated to other devices with fast changing physical processes. From Fig. \ref{example}, we can also see that the proposed QMIX approach achieves better estimation accuracy compared to the sampling policy based on DQN and the uniform sampling policy. This implies that the proposed QMIX approach enables the devices to cooperatively train the sampling policy so as to monitor the physical process accurately.

\begin{figure}[t]
\centering
\setlength{\belowcaptionskip}{-0.45cm}
\vspace{-0.1cm}
\includegraphics[width=8cm]{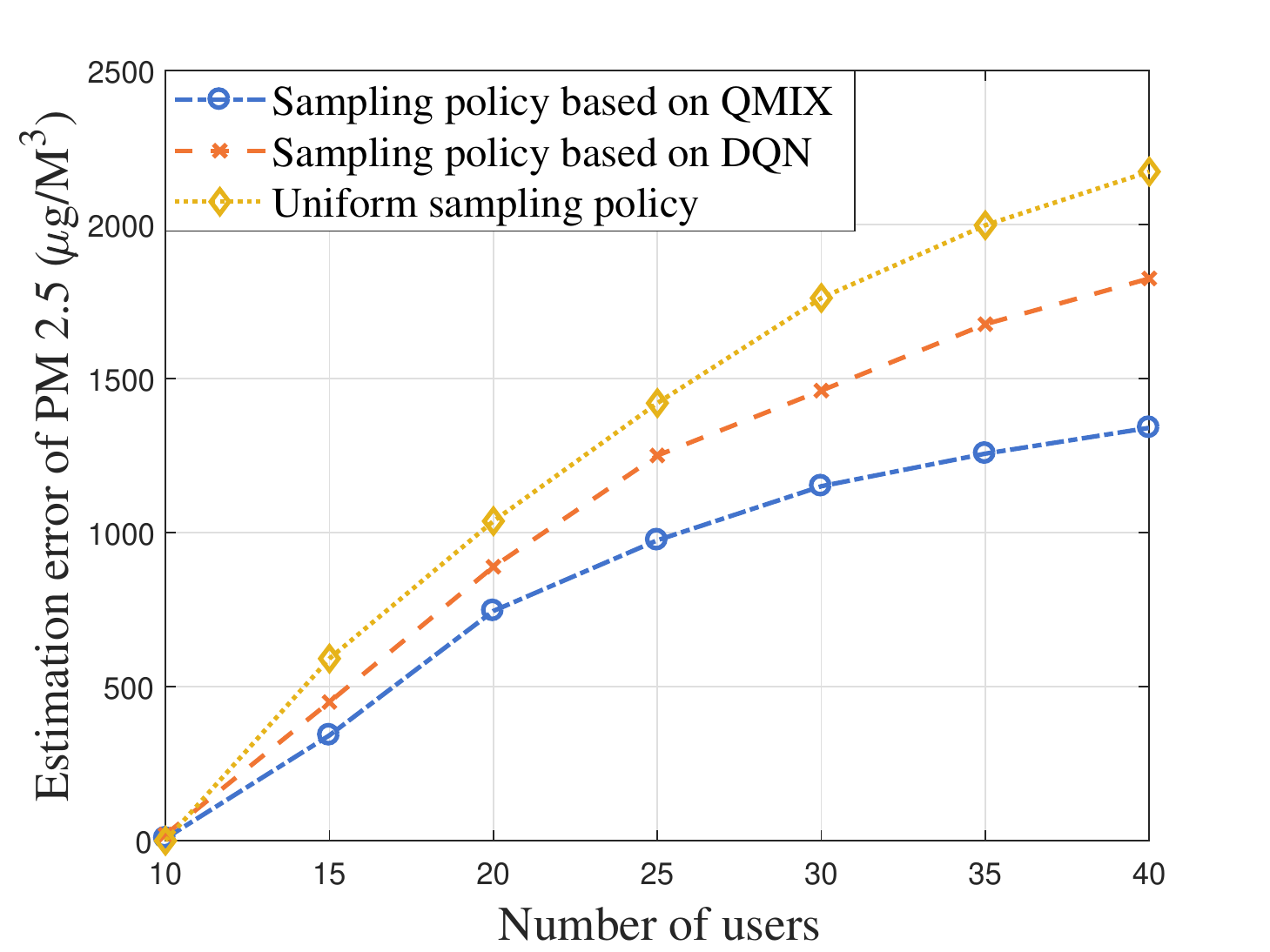}
\centering
\vspace{-0.3cm}
\caption{\small The estimation error of the dynamic physical process as the number of users varies.}
\vspace{-0.3cm}
\label{fig2}
\end{figure}

Fig. \ref{fig2} shows how the estimation error changes as the number of users varies. Clearly, as the number of users increases, the estimation error increases. This is because that the number of RBs is limited in the considered system. Thus, as the number of users increases, the probability that each device can be allocated RBs for status uploading will decrease. In consequence, each device must increase the sample interval to ensure that the sampled packet can be transmit timely, which results in an increase of estimation error. Fig. \ref{fig2} also shows that, as the number of devices continues to increase, the estimation error increases slowly. This is due to the fact that as the number of devices is much larger than the number of RBs, the BS cannot collect the sampled packet from each device immediately and, as a result, the estimation error increases slowly since the outdated sampled packet are ineffective for monitoring physical processes. From Fig. \ref{fig2}, we can also see that the proposed algorithm reduces the estimation error by up to 19.4\% and 33.8\% compared to the sampling policy based on DQN and the uniform sampling policy. This gain stems from the fact that the proposed algorithm enables the devices to cooperatively adjust their sampling policies based on the global state information collected by the BS thus achieving a better monitoring performance.

\begin{figure}[t]
\centering
\setlength{\belowcaptionskip}{-0.45cm}
\vspace{-0.1cm}
\includegraphics[width=8.2cm]{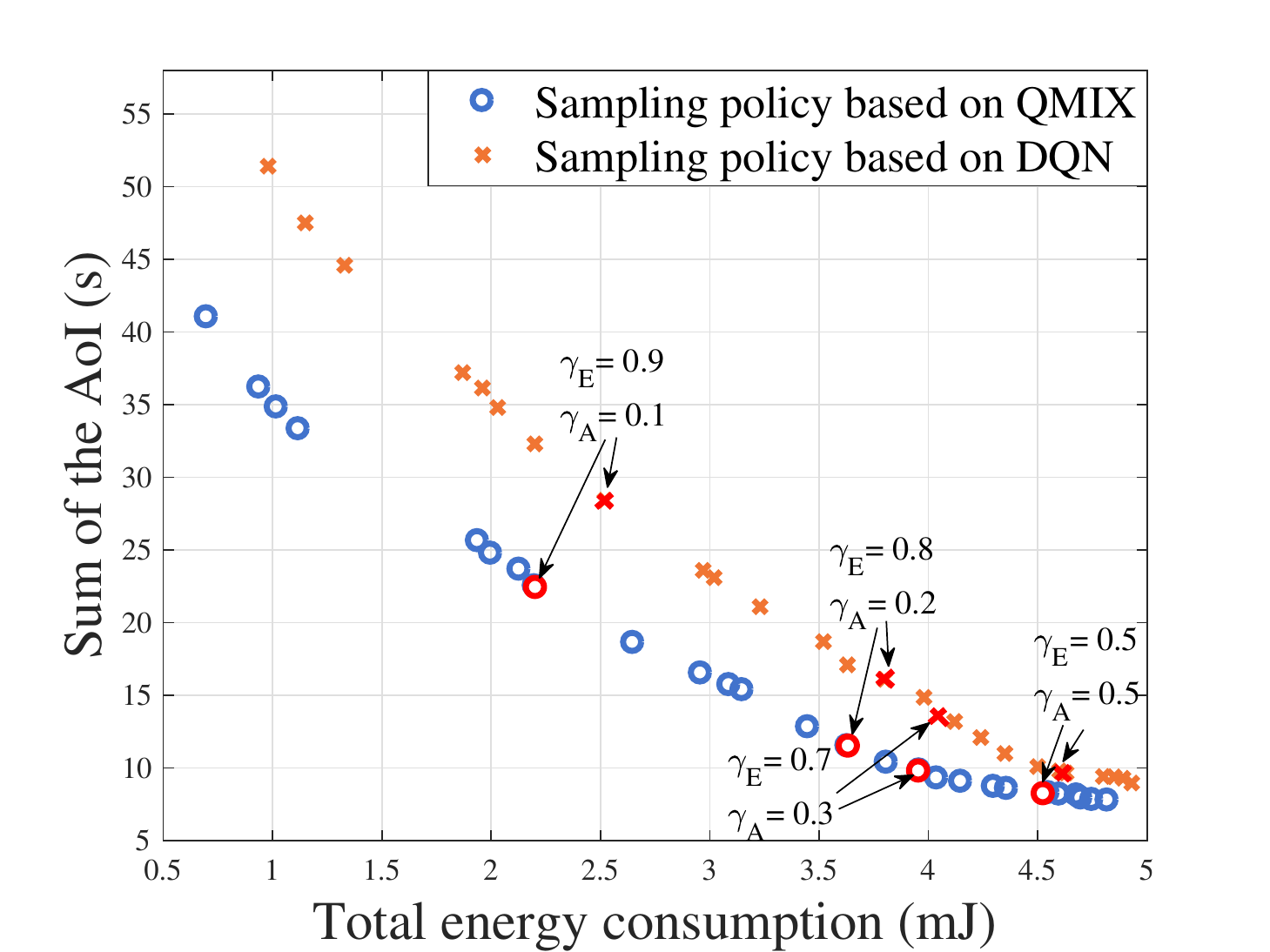}
\centering
\vspace{-0.2cm}
\caption{\small The minimum of the energy consumption and the AoI of all devices change as the scaling parameters vary.}
\vspace{-0.4cm}
\label{Pareto}
\end{figure}

Fig. \ref{Pareto} shows how the minimum of the energy consumption and the AoI of all devices change as the scaling parameters vary. In this figure, each point represents the minimum energy consumption and the AoI achieved by the considered algorithms under the given scaling parameters $\gamma_{\rm A}$ and $\gamma_{\rm E}$. Here, as $\gamma_{\rm E}$ decreases while $\gamma_{\rm A}$ increases, the minimum AoI of all devices decreases. This is because as $\gamma_{\rm E}$ decreases and $\gamma_{\rm A}$ increases, the considered algorithms will focus more on the minimization of the AoI of all devices. Thus, the IoT devices must increase the sampling frequency to capture the variation of the physical process so as to minimize their AoI. In Fig. \ref{Pareto}, we also see that, the proposed QMIX algorithm achieves a better performance. This stems from the fact that, the proposed QMIX algorithm enables the distributed devices to adjust their sampling policies by using the global state information collected by the BS. Fig. \ref{Pareto} also shows that, as $\gamma_{\rm E}$ decreases while $\gamma_{\rm A}$ increases, the gap between the minimum energy consumption and the AoI achieved by DQN and the QMIX approach decreases since the number of RBs that the considered algorithms can optimize is limited.

\vspace{-0.2cm}
\section{Conclusion}
In this paper, we have considered a real-time IoT system to capture the variation of a physical process. We have formulated an optimization problem that seeks to adjust the sampling policy of each distributed device and the device selection scheme of the BS so as to minimize the weighted sum of the AoI and total device energy consumption of all devices. To solve this problem, we have developed a distributed QMIX algorithm that enables edge devices to cooperatively update the RL parameters using the global value function generated by the BS based on the observed global state information thus, improving the performance of the sampling policy. Given the sampling policy, we have optimized the device selection scheme to minimize the weighted sum of AoI and energy consumption of all devices. Simulation results have shown that the proposed approach yields significant gains compared to conventional approaches.

\vspace{-0.2cm}
\section*{Appendix}

\subsection{Proof of Theorem 1}
To determine the optimal device selection scheme $\bm u_t$, we first need to build the relationship between $\bm u_t$ and the sampling policy $\bm s_t$. For this purpose, the AoI of devices in \eqref{eq:AOISUM} can be rewritten as
\begin{equation}\label{eq:SUM}
\begin{aligned}
\!&\gamma_{
\rm A}\overline \Phi_{t}(\bm s_{t},\bm u_{t})+\gamma_{
\rm E}\overline e_{t}(\bm s_{t},\bm u_{t})\\
=&\frac{1}{t}\!\sum\limits_{i = 1}^t \mathbb{E} \left[\sum\limits_{m = 1}^M \!\!\gamma_{ \rm A}\!\left(\Phi_{m,i}(s_{m,i},u_{m,i})\!+\!\gamma_{
\rm E}e_{m,i}(s_{m,i},u_{m,i})\right)\right]\\
=&\frac{1}{t}\!\sum\limits_{i = 1}^t \!\sum\limits_{m = 1}^M\! \mathbb{E}\left[\gamma_{ \rm A}\! \left[(1\!-\!u_{m,i})(\Phi_{m,i-1}\!+\! \tau)\!+\! u_{m,i}(l_{m,i}(u_{m,i})\! \right. \right.\\
& \!\!\qquad \qquad \qquad\left. \left.+\phi_{m,t}(s_{m,i}))\right]\!+\!\gamma_{ \rm E}\!\left[s_{m,i}C_S\!+\!P_{T}l_{m,i}(u_{m,i})\right] \right]\\
=&\frac{1}{t}\!\sum\limits_{i = 1}^t \!\sum\limits_{m = 1}^M \!\!\left[\gamma_{ \rm A}\! (1\!-\!u_{m,i})(\Phi_{m,i\!-\!1}\!\!+\! \tau)\!+\! u_{m,i}(D_{m,i}\!\!+\!\!\phi_{m,i}(s_{m,i})) \right.\\
& \qquad \qquad \qquad \qquad \qquad \qquad \qquad\left.\!+\!\gamma_{ \rm E}\left[s_{m,i}C_S\!+\!P_{T}D_{m,i}\right] \right]
\end{aligned}
\end{equation}
where $D_{m,i}(u_{m,i})=\mathbb E\left[l_{m,i}(u_{m,i})\right]$ and $\Phi_{m,i-1}$ is the simplified notation for $\Phi_{m,i-1}( s_{m,i-1},u_{m,i-1})$. According to (\ref{eq:SUM}), the weighted sum of AoI and energy consumption of device $m$ at time slot $t$ can be written by
\begin{equation}\label{eq:SUMM}
\begin{aligned}
&\!\!\gamma_{\rm A}\Phi_{m,t}(s_{m,t},u_{m,t})+\gamma_{\rm E}e_{m,t}(s_{m,t},u_{m,t})\\
=&\gamma_{\rm A}\! \left[(1\!-\!u_{m,t})(\Phi_{m,t-1}\!+\! \tau)\!+\! u_{m,t}(D_{m,t}\!+\!\phi_{m,t})\right]\\
&\qquad \qquad \qquad \qquad \qquad \qquad \qquad \!+\!\gamma_{ \rm E}\left(s_{m,t}C_S\!+\!P_{T}D_{m,t}\right)\\
=&u_{m,t}\!\left[\gamma_{\rm A}\!\left(D_{m,t}\!\!+\!\!\phi_{m,t}\!-\!\Phi_{m,t\!-\!1}\!\!-\!\! \tau \right)\!+\!\gamma_{\rm E}P_{T}D_{m,t}\right]\\
&\qquad \qquad \qquad \qquad \qquad \qquad\!+\!\gamma_{ \rm E}s_{m,t}C_S\!+\!\gamma_{ \rm A}(\Phi_{m,t-1}\!+\!\tau)  \\
=&C_{m,t,1} u_{m,t}\!+\!C_{m,t,2},
\end{aligned}
\end{equation}
where $D_{m,t}$ and $\phi_{m,t}$ are short for $D_{m,t}(\!u_{m,t})$ and $\phi_{m,t}(\!s_{m,t})$, respectively. $C_{m,t,1}\!\!=\!\gamma_{\rm A}\!\left(D_{m,t}(1)\!+\!\phi_{m,t}(s_{m,t})\!-\!\!\Phi_{m,t-1}\!-\! \tau \right)\!+\!\gamma_{\rm E}P_{T}D_{m,t}(u_{m,t})$ and $C_{m,t,2}=\gamma_{ \rm E}s_{m,t}C_S+\!\gamma_{ \rm A}(\Phi_{m,t-1}\!+\!\tau)$\\
with $\phi_{m,t}(s_{m,t})\!=\!s_{m,t}\max\{0,\delta(t)\!-\!\Delta_{m,t}\}\!+\!(1\!-\!s_{m,t}){\rm min}\{\phi_{m,t-1}\!+\! \tau,\!\phi_{\rm max}\})$.

Given the sampling policy $\bm s_t$, the device selection problem at time $t$ can be written as
\begin{subequations}
\begin{align}
\mathop{\min }_{u_{m,t}}\quad
		& \sum_{m=1}^M (C_{m,t,1} u_{m,t}+C_{m,t,2})\tag{\theequation}\\
		\textrm{s.t.}\quad\:\:
		& \sum\limits_{m \in \mathcal{M}}  u_{m,t} \leqslant I,\\
& u_{m,t}\in\{0,1\}, \quad\forall m \in\mathcal M.
\end{align}
\end{subequations}

Denote the net set $\mathcal M_1=\{m\in\mathcal M|C_{m,t,1}<0\}$.
If $|\mathcal M_1|\leqslant I$, we have
\begin{equation}
\begin{aligned}
\!\!&u_{m,t}^*
=\left\{ \begin{array}{l}
\!\!1,\quad\;{\rm if} \; m \in \mathcal M_1\\
\!\!0 ,\quad\;{\rm otherwise}.
\end{array} \right.\
\end{aligned}
\end{equation}

\subsection{Proof of Lemma 1}
In order to prove Lemma 1, we first need to build the relationship between the Bellman operator ${\rm H}Q$ in traditional Q network over global observable information and the Bellman operator ${\rm H}Q_{\rm tot}$ in QMIX over partially observable information. As shown in \cite{WMD}, the Bellman operator in traditional Q network over global observable information is defined as $({\rm H}Q)(\bm o_t,\bm a_t)
\!=\!\sum\limits_{\bm o_t'}\! P_{\bm a_t}(\bm o_t,\bm o_t')\!\left[ R(\bm o_t,\bm a_t)\!+\!\gamma  \mathop {\max }\limits_{\bm a_t' \in \mathcal{A}} \left( Q^*(\bm o_t', \bm a_t') \right)\right]$. Using Definition 1, the Bellman operator in QMIX method can be expressed as
\begin{equation}\label{eq:optimaltransferstatevalue}
\begin{aligned}
&\!\!Q^*_{\rm tot}(\bm o_t, \bm a_t)\\
=&Q^*(\bm o_t, \bm a_t)\!-\!\varepsilon(\bm o_t, \bm a_t)\\
=&\!\sum\limits_{\bm o_t'}\!\! P_{\bm a_t}(\bm o_t,\bm o_t')\!\left[ R(\bm o_t,\bm a_t)\!+\!\gamma \mathop {\max }\limits_{\bm a_t' \in \mathcal{A}} \left( Q^*(\bm o_t', \bm a_t') \right) \right]\!\!-\!\varepsilon(\bm o_t, \bm a_t)\\
=&\!\!\sum\limits_{\bm o'}\!\! P_{\bm a_t}(\bm o_t,\!\bm o_t')\!\!\left[\!R(\bm o_t,\bm a_t)\!+\!\gamma\!\!\left(\!\!f\!\!\left(\!\!\mathop {\max }\limits_{\bm a'_{1,t} \in \mathcal{A}}\!\!Q_1^*(\bm o'_{1,t}, \bm a'_{1,t}),\ldots\!,\right.\right.\right. \\
&\qquad \quad \qquad \left. \left. \left.\!\!\!\mathop {\max }\limits_{\bm a_{M,t}' \in \mathcal{A}}\!\!Q_M^*(\bm o'_{M,t}, \bm a'_{M,t})\!\!\right)\!\!+\!\varepsilon(\bm o_t',\!\bm a_t')\!\!\right)\! \!-\!\varepsilon(\bm o_t,\!\bm a_t)\!\right]\!\!,
\end{aligned}
\end{equation}
where the last equation follows from the fact that the non-negativity
of the mixing network weights and the availability of $Q_m(\bm o_{m,t},\!s_{m,t})$ of each device $m$ at each time slot (i.e., $u_{m,t}\!=\!1$). Thus, the optimal QMIX function is defined for a generic function $Q_{\rm tot}\!:\!\{\mathcal{O_{\rm 1},\cdots,\!O_{\rm M}}\} \!\times\! \{\mathcal{A_{\rm 1},\cdots,\!A_{\rm M}}\} \! \rightarrow\! \mathbb{R}$ as
\begin{equation}\label{eq:contraction}
\begin{aligned}
&\!\!({\rm H} Q_{\rm tot})(\bm o_t,\bm a_t)\!\!=\!\!\sum\limits_{\bm o_t'}\! P_{\bm a_t}(\bm o_t,\!\bm o_t')\!\bigg [\! R(\bm o_t,\!\bm a_t)-\varepsilon(\bm o_t,\!\bm a_t)  \\
&\left.\!\!\!\!\!+\gamma\!\!\left( \!\!f\!\!\left(\!\mathop {\max }\limits_{\bm a'_{1,t} \in \mathcal{A}}\!\!\!Q_1^*(\bm o'_{1,t},\!\bm a'_{1,t}),\! \ldots,\!\!\!\!\mathop {\max }\limits_{\bm a_{M,t}' \in \mathcal{A}}\!\!\!Q_M^*\!(\bm o'_{M,t},\! \bm a'_{M,t})\!\!\right)\!\!\!+\!\varepsilon(\bm o_t',\!\bm a_t')\!\!\right)\!\! \right]\!\!.
\end{aligned}
\end{equation}
Next, we prove that ${\rm H} Q_{\rm tot}$ is a contraction in the sup-norm. Based on (28), we have:
\begin{equation}\label{eq:contractionproof}
\begin{aligned}
&\Vert {\rm H}Q_{\rm tot}^1-{\rm H}Q_{\rm tot}^2 \Vert_\infty \\
=&\mathop{\max }\limits_{\bm o_t,\bm a_t} \left|\sum\limits_{\bm o'_t}\!\! P_{\bm a}(\bm o_t,\bm o'_t) \gamma \left(Q_{\rm tot}^1(\bm o_t', \bm a_t') -Q_{\rm tot}^2(\bm o_t', \bm a_t') \right)\right| \\
=&\mathop{\max }\limits_{\bm o_t,\bm a_t}\! \sum\limits_{\bm o_t'}\!P_{\bm a_t}(\bm o_t,\bm o_t') \left| \gamma \mathop {\max }\limits_{\bm a_t' \in \mathcal{A}} \left( Q^1_{\rm tot}(\bm o_t', \bm a_t') - Q^2_{\rm tot}(\bm o_t', \bm a_t')\right)\right|\\
\leqslant &\mathop{\max }\limits_{\bm o_t,\bm a_t}\!\gamma\!\sum\limits_{\bm o_t'}\!\!P_{\bm a_t}(\bm o_t,\bm o_t')\mathop{\max }\limits_{\bm o_t',\bm a_t'}\left| Q_{\rm tot}^1(\bm o_t',\bm a_t')-Q_{\rm tot}^2(\bm o_t',\bm a_t')\right| \\
=&\mathop{\max }\limits_{\bm o_t,\bm a_t}\gamma\!\sum\limits_{\bm o_t'}\!P_{\bm a_t}(\bm o_t,\bm o_t')\Vert Q_{\rm tot}^1-Q_{\rm tot}^2\Vert_\infty \\
=& \gamma \Vert Q_{\rm tot}^1-Q_{\rm tot}^2\Vert_\infty
\end{aligned}
\end{equation}
This completes the proof.

\subsection{Proof of Theorem 2}
Using the definition of ${\rm H} Q_{\rm tot}$ in Lemma 1, the update rule of QMIX method that introduces $\varepsilon(\bm o_t, \bm a_t)$ can be written by
\begin{equation}\label{eq:update}
\begin{aligned}
\!\!\!&Q_{\rm tot}^{i+1}(\bm o_t, \bm a_t)\! \leftarrow\! (1\!-\!\alpha)Q_{\rm tot}^{i}(\bm o_t, \bm a_t)\\
&\!+\!\alpha\!\left(\!\! R (\bm o_t, \bm a_t)\!+\!\gamma \mathop {\max }\limits_{\bm a_t' \in \mathcal{A}}\left(Q_{\rm tot}^{i}(\bm o_t', \bm a_t')\!+\!\varepsilon(\bm o_t', \bm a_t')\right)\!-\!\varepsilon(\bm o_t, \bm a_t)\!\!\right)\!\!.
\end{aligned}
\end{equation}
Subtracting $Q^*_{\rm tot}(\bm o_t, \bm a_t)$ from both sides in (\ref{eq:update}) and letting $\Lambda^{i} =Q^*_{\rm tot}(\bm o_t, \bm a_t)-Q_{\rm tot}^{i}(\bm o_t, \bm a_t)$, we have
\begin{equation}\label{eq:updateerror1}
\begin{aligned}
&\Lambda^{i+1} \leftarrow (1\!-\!\alpha)\Lambda^{i}\!+\!\alpha F^{i}(\bm o_t, \bm a_t)
\end{aligned}
\end{equation}
where $F^{i}(\bm o_t,\!\bm a_t)\!=\!R (\bm o_t, \bm a_t)\!+\!\gamma\! \mathop {\max }\limits_{\bm a' \in \mathcal{A}}\left(Q_{\rm tot}^{i}(\bm o_t', \bm a'_t)\!+\!\varepsilon(\bm o',\!\bm a')\!\right)$\\$\!-\varepsilon(\bm o_t, \bm a_t)\!-\!Q^*_{\rm tot}(\bm o_t, \bm a_t)$. Based on \cite{Shou}, the random process $\{\!\Lambda^{i}\!\}$ converges to 0 under the following conditions: 

a) $\mathbb{E}\left[ F^{i}(\bm o_t, \bm a_t)| \mathcal{F}^{i}\right]\!\!\leqslant\!\gamma \Vert \Lambda^{i} \Vert_\infty$; 

b) ${\rm var}\left[ F^{i}(\bm o_t, \bm a_t)| \mathcal{F}^{i}\right]\leqslant C(1+\Vert \Lambda^{i} \Vert_\infty^2)$. 

Next, we prove that the random process $\{\Lambda^{i}\}$ in QMIX method satisfies a) and b), respectively.

For a), we have
\begin{equation}\label{eq:errordes}
\begin{aligned}
\!\!&\mathbb{E}\!\left[ F^{i}(\bm o_t, \bm a_t)| \mathcal{F}^{i}\right]\!\!\\
=&\sum\limits_{\bm o'}\! P_{\bm a}(\bm o_t,\!\bm o_t')\!\!\left[\!R (\bm o_t,\!\bm a_t)\!+\!\gamma \mathop {\max }\limits_{\bm a' \in \mathcal{A}}\left(Q_{\rm tot}^{i}(\bm o_t',\!\bm a_t')\!\!+\!\varepsilon(\bm o_t',\!\bm a_t')\right)\!\right.\\
&\qquad\qquad\qquad\qquad\qquad\qquad\qquad-\!\varepsilon(\bm o_t,\!\bm a_t)\!-\!Q^*_{\rm tot}(\bm o_t,\!\bm a_t)\!\bigg] \\
=&({\rm H}Q_{\rm tot})(\bm o_t,\!\bm a_t)\!-\!({\rm H}Q_{\rm tot}^*)(\bm o_t,\!\bm a_t)\\
\leqslant & \gamma \Vert Q_{\rm tot}-Q^*_{\rm tot}\Vert_\infty \\
=&\gamma \Vert \Lambda \Vert_\infty,
\end{aligned}
\end{equation}
where the last equation stems from the fact $Q_{\rm tot}^*={\rm H}Q_{\rm tot}^*$ and the last inequality follows from Lemma 1.

For b), we have
\vspace{-0.2cm}
\begin{equation}\label{eq:errorvar}
\begin{aligned}
&\!{\rm var}\left[ F^{i}(\bm o_t, \bm a_t)| \mathcal{F}^{i}\right] \\
=& \mathbb{E}\!\!\left[\! \left( \!\!R (\bm o_t,\!\bm a_t)\!\!+\!\gamma\!\mathop {\max }\limits_{\bm a_t' \in \mathcal{A}}\!\left(Q_{\rm tot}^{i}(\bm o_t',\! \bm a_t')\!+\!\varepsilon(\bm o_t',\!\bm a_t')\right)\!-\!\varepsilon(\bm o_t,\!\bm a_t)\! \right. \right. \\
&\left. \left. \;\;\qquad \qquad-Q^*_{\rm tot}(\bm o_t,\!\bm a_t)\!-\!({\rm H}Q_{\rm tot})(\bm o_t,\!\bm a_t)\!+\!Q_{\rm tot}^*(\bm o_t,\!\bm a_t)\!\right)^{\!2} \right]\\
=&\mathbb{E}\!\!\left[\! \left(\! R (\bm o_t, \bm a_t)\!+\!\gamma \mathop {\max }\limits_{\bm a' \in \mathcal{A}}\left(Q_{\rm tot}^{i}(\bm o_t', \bm a_t')\!+\!\varepsilon(\bm o_t', \bm a_t')\right)\!-\!\varepsilon(\bm o_t, \bm a_t)\! \right. \right. \\
&\left. \left. \quad\qquad\qquad\qquad\qquad\qquad\;\;\;\qquad \qquad -\!({\rm H}Q_{\rm tot})(\bm o_t,\bm a_t)\!\right)^2 \right]\\
=&{\rm var}\!\!\left[\! R (\bm o_t,\!\bm a_t)\!+\!\!\gamma \mathop {\max }\limits_{\bm a_t' \in \mathcal{A}}\!\!\left(Q_{\rm tot}^{i}(\bm o_t',\!\bm a_t')\!+\!\varepsilon(\bm o_t',\!\bm a_t')\right)\!\!-\!| \mathcal{F}\right]\\
\leqslant & C(1+\Vert \Lambda^{i} \Vert_\infty^2),
\end{aligned}
\end{equation}
where $C$ is a constant and the last inequality stems from the fact that $R (\bm o_t, \bm a_t)$, $\varepsilon(\bm o_t, \bm a_t)$, and $\varepsilon(\bm o_t', \bm a_t')$ are bounded. From (\ref{eq:errordes}) and (\ref{eq:errorvar}), we can see that, the proposed QMIX algorithm satisfies the conditions in a) and b), hence, $Q_{\rm tot}(\bm o_t, \bm a_t)$ reaches $Q^*_{\rm tot}(\bm o_t, \bm a_t)$ as $\Lambda^{i}=0$. This completes the proof.

\bibliographystyle{IEEEbib}
\def\baselinestretch{1}
\bibliography{AOI}
\end{document}